\documentclass{emulateapj}
\usepackage{placeins}
\usepackage{graphicx}

\shorttitle{Protostellar disk accretion} \shortauthors{Zhu et al.}

\begin{document}

\title{Long-term Evolution of Protostellar and Protoplanetary Disks. II. Layered Accretion with Infall}

\author{Zhaohuan Zhu\altaffilmark{1}, Lee Hartmann\altaffilmark{1}, and
Charles Gammie \altaffilmark{2,3}}

\altaffiltext{1}{Dept. of Astronomy, University of Michigan, 500
Church St., Ann Arbor, MI 48109} \altaffiltext{2}{Dept. of
Astronomy, University of Illinois Urbana-Champaign, 1002 W. Green
St., Urbana, IL 61801} \altaffiltext{3}{Dept. of Physics, University
of Illinois Urbana-Champaign}

\email{zhuzh@umich.edu, lhartm@umich.edu, gammie@illinois.edu}

\newcommand\msun{\rm M_{\odot}}
\newcommand\lsun{\rm L_{\odot}}
\newcommand\msunyr{\rm M_{\odot}\,yr^{-1}}
\newcommand\be{\begin{equation}}
\newcommand\en{\end{equation}}
\newcommand\cm{\rm cm}
\newcommand\kms{\rm{\, km \, s^{-1}}}
\newcommand\K{\rm K}
\newcommand\etal{{\rm et al}.\ }
\newcommand\sd{\partial}
\newcommand\mdot{\rm \dot{M}}
\newcommand\rsun{\rm R_{\odot}}
\newcommand\yr{\rm yr}

\begin{abstract}

We use one-dimensional two-zone time-dependent accretion disk models
to study the long-term evolution of protostellar disks subject to
mass addition from the collapse of a rotating cloud core. Our model
consists of a constant surface density magnetically coupled active
layer, with transport and dissipation in inactive regions only via
gravitational instability.  We start our simulations after a central
protostar has formed, containing $\sim$ 10\% of the mass of the
protostellar cloud. Subsequent evolution depends on the angular
momentum of the accreting envelope.  We find that disk accretion
matches the infall rate early in the disk evolution because much of
the inner disk is hot enough to couple to the magnetic field. Later
infall reaches the disk beyond $\sim$10 AU, and the disk undergoes
outbursts of accretion in FU Ori-like events as described in Zhu et
al. 2009c.  If the initial cloud core is moderately rotating most of
the central star's mass is built up by these outburst events.  Our
results suggest that the protostellar ``luminosity problem'' is
eased by accretion during these FU Ori-like outbursts.  After infall
stops the disk enters the T Tauri phase.  An outer, viscously
evolving disk has structure that is in reasonable agreement with
recent submillimeter studies and its surface density evolves from
$\Sigma \propto R^{-1}$ to $R^{-1.5}$. An inner, massive belt of
material-- the ``dead zone'' -- would not have been observed yet but
should be seen in future high angular resolution observations by
EVLA and ALMA. This high surface density belt is a generic
consequence of low angular momentum transport efficiency at radii
where the disk is magnetically decoupled, and would strongly affect
planet formation and migration.

\end{abstract}


\section{Introduction}
In the picture of low-mass star formation from large, cold
protostellar clouds, any small initial rotation of the cloud will
lead to the formation of an accretion disk to conserve angular
momentum; thus much, if not most, of the mass of low-mass stars is
probably accreted from disks. However, if disks transport the
infalling mass steadily to the central star, accretion luminosities
are considerably higher than typically observed protostellar
luminosities (Kenyon \etal 1990, 1994; Enoch \etal 2008; Evans \etal
2009). This ``luminosity problem'' can be alleviated if disks spend
most of their evolution accreting slowly, undergoing episodic
accretion outbursts where most of the mass is added to the central
stars.  One implication of non-steady accretion is that mass may
pile up in specific regions of disks, which could have significant
consequences for understanding planet formation.

Over the past decade, disk angular momentum transport by the
magnetorotational instability (MRI) (Balbus \& Hawley 1998) and by
the gravitational instability (GI) (Durisen \etal 2007) has become
much better understood.  It seems likely that both types of
instability need to be considered to understand protostellar
accretion.  At early evolutionary stages the disk is likely to be
quite massive with respect to the central protostar, suggesting that
GI may be important; in addition, these disks are so cold that
thermal ionization cannot sustain the MRI, which requires coupling
of the magnetic field to the mostly-neutral gas through collisions
(e.g., Reyes-Ruiz $\&$ Stepinski 1995). On the other hand, external
ionizing sources (cosmic and/or X rays) can provide the necessary
ionization for MRI activitation up to a limiting surface density,
ionizing the outer surfaces of the disk and leading to accretion in
an ``active layer'', leaving a ``dead zone'' in the midplane (Gammie
1996). In addition, at high accretion rates, the inner disk can
become thermally-ionized, as in FU Ori outbursts (Zhu \etal 2007,
2009b).

Mismatches in the transport rate between the GI and MRI can lead to
outbursts of accretion (Armitage et al. 2001, Zhu et al. 2009 bc).
In Zhu et al. 2010 (Paper I), we constructed one-dimensional,
time-dependent disk models to study the protostellar unsteady
accretion. These models assume both thermally-activated and layered
MRI-driven accretion along with a local treatment of the GI.  We
further adopted steady mass addition at a specified outer disk
radius to drive the system. We found that accretion in protostellar
disks is unsteady over a wide range of parameters because of a
mismatch between GI transport in the outer disk and MRI transport in
the inner disk. Our results for outburst behavior in these
one-dimensional model were sufficiently comparable to our previous
two-dimensional simulations of outbursts (Zhu \etal 2009c),
confirming the utility of the faster 1D simulations to explore wider
ranges of parameter space.

In this paper we extend our results from Paper I to study long-term
disk evolution from the protostellar phase to the T Tauri phase,
using a more self-consistent treatment for infall from a collapsing,
rotating cloud core. This allows us to study the effect of the
initial core rotation and different disk accretion configurations
(layered accretion, GI-only accretion, and constant-$\alpha$
accretion) on the disk formation and evolution.  Rice, Mayo, \&
Armitage (2009) have also investigated somewhat similar
one-dimensional models but do not assume layered accretion.

Our models treat the same phase of evolution as Vorobyov \& Basu (2005,
2006, 2007) investigated with two-dimensional models.  They also argue
that protostellar accretion is generally non-steady, but for a different
reason, specifically the accretion of clumps created by gravitational
instabilities.  Kratter \etal (2009) have also investigated the
protostellar accretion phase with three-dimensional numerical simulations.
Although our treatment of the core collapse and the GI is much more
schematic than used in the above investigations, we are able to treat
radiative cooling more realistically, and can follow disk evolution to
much smaller radii, where the (thermally-activated) MRI can become
important. Furthermore, our two-zone disk model allows us to study the
effects of different disk accretion configurations efficiently. The
variety of disk structures predicted by our simulations with different
initial cloud core rotations and disk accretion mechanisms can be tested
by future EVLA and ALMA observations, which will help us to better
understand disk accretion and planet formation.

We describe our one-dimensional, two-zone model with self-consistent
infall in \S 2.  In \S 3 we explore the behavior of the disk models
with infall.  In \S 4 we discuss the implications of our results,
and summarize our conclusions in \S 5. We defer all derivations to
the Appendix.

\section{1D2Z models with infall}
The one-dimensional two zone (1D2Z) disk model has been introduced
in Paper I. It consists of two vertically averaged layers (the
surface layer and the ``dead zone'') evolving independently.  The
surface density evolves according to the mass and angular momentum
conservation equations in cylindrical coordinates,
\begin{equation}
2\pi R\frac{\partial \Sigma_{i}}{\partial t}-\frac{\partial
\dot{M_{i}}}{\partial R}=2\pi g_{i}(R,t)\,,\label{eq:1}
\end{equation}
\begin{equation}
2\pi R\frac{\partial}{\partial t}(\Sigma_{i} R^{2}
\Omega)-\frac{\partial}{\partial
R}(\dot{M_{i}}R^{2}\Omega)=2\pi\frac{\partial}{\partial
R}(R^{2}W_{R\phi,i})+2\pi\Lambda_{i}\,,\label{eq:2}
\end{equation}
where $\Sigma$ is the total surface density, $\Omega$
is the angular frequency, $\dot{M_{i}}=-2\pi\Sigma_{i} R v_{i}$ is
the radial mass flux in the disk, $W_{R\phi,i}=R\Sigma_{i}
\nu_{i}d\Omega/dR$, $\nu_{i}$ is the viscosity, and $i$ denotes `a'
(active layer) or `d' (dead zone). 2$\pi$$g_{i}(R,t)$ and
The terms 2$\pi$$\Lambda_{i}$ are the mass and angular momentum flux brought
by the infalling matter into the disk (Cassen \& Moosman 1981).
Assuming instantaneous centrifugal balance,
\footnote{The validity of this assumption with infall has
been shown in Appendix B in Cassen \& Moosman (1981).
Instantaneous centrifugal balance is a good approximation in
our simulations because the change of the central star mass
on an orbital time is very small.}
equations \ref{eq:1} and \ref{eq:2} can be simplified
to
\begin{eqnarray}
\dot{M_{i}}&=&6\pi R^{1/2}\frac{\partial}{\partial
R}(R^{1/2}\Sigma_{i}\nu_{i})+\frac{2\pi
R^{2}\Sigma_{i}}{M_{r}}\frac{\partial M_{r}}{\partial t}\nonumber\\
&&-4\pi(\frac{R}{GM_{r}})^{1/2}(\Lambda_{i}-g(R,t)R^{2}\Omega(R))\,,\label{eq:3}
\end{eqnarray}
where $M_{r}$ is the sum of the mass of the central star (M$_{c}$) and
the disk mass within R, using the approximation of the
gravitational potential adopted by Lin $\&$ Pringle (1990). The first
term in equation \ref{eq:3} represents disk accretion due to
viscosity; the second term is due to the central star mass changing
with time; and the third term is due to the infalling matter (Cassen
\& Moosman 1981). Since the infalling matter only falls onto the
surface of the disk, we assume $g_{d}$ and $\Lambda_{d}$ are 0, thus
$g_{a}$ and $\Lambda_{a}$ can be written as $g$ and $\Lambda$ for
short. The effect of the infalling matter onto the disk is limited
to just two free functions: g(R,t) and $\Lambda(R,t)$.

The surface densities $\Sigma_{a}$ and $\Sigma_{d}$ exchange mass as
described in Paper I to maintain $\Sigma_{a}\leq\Sigma_{A}$, where
$\Sigma_{A}$ is the maximum MRI active layer surface density and is
assumed to be a constant throughout the disk (Gammie 1996). We solve
equation \ref{eq:3} and \ref{eq:1} sequentially: $\dot{M_{i}}$ is
calculated with the current disk temperature and $\Sigma_{i}$, and
then it is inserted into equation \ref{eq:1} to update $\Sigma_{i}$
at the next timestep.

The disk temperature is determined by the balance between the
heating and radiative cooling. Here the external temperature
$T_{ext}$, which represents the heating effect of the irradiation
from the central star is assumed to be
\begin{equation}
T_{ext}^{4}=\frac{fL}{4\pi R^{2}\sigma}\,,\label{eq:text}
\end{equation}
where $L$ is the total luminosity of the star and changes with the
central star mass by
\begin{equation}
\frac{L_{*}}{L_{\odot}}=\frac{8M_{*}}{9M_{\odot}}+\frac{1}{9}
\end{equation}
which is a fit to the T-Tauri star birthline from Kenyon \&
Hartmann(1995). The change of luminosity  has little effect on the
disk evolution in our simulations. $\sigma$ is the Stefan-Boltzmann
constant, and $f$ is the coefficient, which accounts for the
non-normal irradiation from the central star at the disk's surface.
We assume $f$=0.1 in this paper. The viscous heating rate of the
disk is
\begin{equation}
Q_{visc}=\frac{3}{2}W_{R\phi}\Omega\,,
\end{equation}
where the stress $W_{R\phi}=3/2\Sigma_i \nu_{i}\Omega$ assuming
Keplerian rotation. Both gravitational and magneto-rotational
instabilities are considered. The anomalous viscosity $\nu_i$ is
\begin{equation}
\nu_i=\alpha_i\frac{c_{s_i}^{2}}{\Omega}\,,\label{eq:nu}
\end{equation}
where $c_{s_{i}}$ are the sound speed of the active layer and the
dead zone, $\alpha_i=\alpha_{Q}+\alpha_{M}$ and
\begin{equation}
\alpha_{Q}=e^{-Q^{4}}\,.
\end{equation}
The Toomre parameter $Q$ is calculated using the disk central
temperature (T$_{d}$) and the total surface density
($\Sigma_{a}$+$\Sigma_{d}$). The form of $\alpha_Q$ is motivated by
a desire to make gravitational torques become important only when $Q
\lesssim 1.4$, as indicated by global three-dimensional simulations
(e.g., Boley et al 2006).  The parameter $\alpha_{M}$ is set to be
0.01 when the disk temperature is above the preset critical MRI
temperature for thermal activitation (T$_{M}$=1500 K is assumed in
this paper) or at the surface active layer. Since the rise time of
MRI is about the orbital timescale  (Stone \etal 1996), we assume
that the MRI viscosity grows on the orbital timescale up to
$\alpha_{M}$ whenever the temperature is above the critical
temperature (T$_{M}$). The effects of different values of
$\alpha_{M}$, T$_{M}$, and $\Sigma_{a}$ on outbursts are discussed
in Paper I. The treatment of radiative cooling is the same as in
Paper I.

Recent observations suggest that the density structure of
protostellar molecular cores, when circularly-averaged, can be
approximated by static Bonnor-Ebert spheres (Alves \etal 2001; di
Francesco et al. 2007; Ward-Thompson et al. 2007).  Though real
clouds have more complex structure (e.g., Benson \& Myers 1989),
resulting in more complicated patterns of infall (Kratter \etal
2009), we adopt this simplified structure to limit the number of
model parameters.  This leads us to use an approximate model of
infall corresponding to that expected from a critical Bonnor-Ebert
sphere, which exhibits similarities to that of the collapse of the
singular isothermal sphere (Terebey, Shu, \& Cassen 1984).

Analytical and numerical
studies ( Foster \& Chevalier 1992; Henriksen et al. 1997; Gong \&
Ostriker 2009) have shown that the collapse of the critical Bonnor-Ebert
sphere can be divided into two stages. In the first stage, the
roughly constant-density inner region collapses to the central
regions nearly simultaneously; then in the second stage, infall from
the remaining cloud core ($\rho\propto r^{-2}$ region \footnote{In
this paper, R denotes the cylindrical radius in the disk while r
denotes the radius in the cloud core.}) is at a rate similar to
c$_{sc}^{3}$/G (c$_{sc}$ is the sound speed at the cloud core
temperature) of the singular isotheral sphere (SIS) collapse model
(Shu 1977). We assume that our simulation starts at the end of the
first stage of very rapid collapse of low-angular momentum material,
so the central star initially has the mass of the inner flat region
of the Bonnor-Ebert sphere, and the collapse of the remaining cloud follows
the singular isothermal sphere solution (Shu 1977).

We use a two-component density profile to represent the Bonnor-Ebert sphere as
in Henriksen et al. (1997). The two density components are
\begin{equation}
\rho=\rho_{c}\,\rm{at}\,\xi<\sqrt{2}\,,
\end{equation}
\begin{equation}
\rho=2\rho_{c}\xi^{-2}\,\rm{at}\,\sqrt{2}<\xi<6.5\,,\label{eq:rho}
\end{equation}
where $\rho_{c}$ is the central density, r is in the normalized unit
$\xi=r/(c_{sc}^{2}/4\pi G \rho_{c})^{1/2}$, $c_{sc}$ is the sound
speed of the cloud and $\rho=2\rho_{c}\xi^{-2}$ is the density
profile of the SIS model. $\xi$=6.5 corresponds to the critical Bonnor-Ebert
sphere radius and the radius of the flat density region
$\xi=\sqrt{2}$ corresponds to $r_{ic}$=$(c_{sc}^{2}/2\pi G
\rho_{c})^{1/2}$.  The total mass of each component is
\begin{equation}
M(r<r_{ic})=\frac{0.27 c_{sc}^{3}}{G^{3/2}\rho_{c}^{1/2}}
\end{equation}
\begin{equation}
M(r_{ic}<r<r_{cr})=\frac{2.88 c_{sc}^{3}}{G^{3/2}\rho_{c}^{1/2}}\,.
\end{equation}
The total mass with this simplified density profile is
$3.1c_{sc}^{3}/G^{3/2}\rho_{c}^{1/2}$, which is close to the mass of
a real critical Bonnor-Ebert sphere $4.2c_{sc}^{3}/G^{3/2}\rho_{c}^{1/2}$.
Since the flat inner region has 10\% mass of the $r^{-2}$ region, we
assume the flat region has 0.1 $M_{\odot}$ and the rest of the Bonnor-Ebert
sphere has 1 $M_{\odot}$.

The rest of the 1 $M_{\odot}$ cloud core beyond $r_{ic}$ collapses
in a manner similar
to that of the singular isothermal sphere, in which case the inner regions
 collapse first due to their shortest free-fall
times and this free-fall zone extends outwards linearly with time
(m$_{0}$c$_{sc}$t, where $m_{0}$=0.975.) (Shu 1977; Terebey, Shu, \&
Cassen 1984). Thus, the mass infall rate is
$\dot{M_{in}}$=m$_{0}$c$_{sc}^{3}$/G, which is constant during the
collapse. At time t, all the material collapsing to the center was
originally at the cloud radius
r$_{0}$=(m$_{0}$/2)c$_{sc}$t+r$_{ic}$, where the addition of
r$_{ic}$ (the radius of the flat density region in the Bonnor-Ebert
sphere) is due to the fact that we define t=0 as the time when the
flat region in the Bonnor-Ebert sphere has already collapsed to the
central star. If the protostellar cloud core is initially in uniform
rotation with angular velocity $\Omega_{c}$, the material falling in
from different directions will have different angular momenta and
arrive at the midplane at differing radii within the so called
``centrifugal radius'' (Cassen \& Moosman 1981)
\begin{equation}
R_{c}=r_{0}^{4}\frac{\Omega^{2}}{GM_{c}}\,,\label{eq:centri}
\end{equation}
where M$_{c}$ is the central star mass. Assuming M$_{c}$ increases
with t , $R_{c} \propto t^{3}$, similar to that found by Terebey,
Shu, \& Cassen (1984) for the singular isothermal sphere collapse.

We use the Cassen \& Moosman (1981) solution for the infall to the
disk at a given time from a spherical cloud initially in uniform
rotation. We start the infall from the transition radius $r_{ic}$ at
$t=0$. Then the mass flux $g(R,t)$ and the angular momentum
$\Lambda$ brought by the infalling matter for a uniformly-rotating
singular isothermal sphere is,
\begin{equation}
g(R,t)=\frac{\dot{M_{in}}}{4\pi
R_{c}}\left(1-\frac{R}{R_{c}}\right)^{-1/2}\,,\label{eq:infallmass}
\end{equation}
\begin{equation}
\Lambda(R,t)=g(R,t)R\left(\frac{GM_{c}}{R_{c}}\right)^{1/2}\,\,
\rm{if} \,\,R\leq R_{c}\,,
\end{equation}
 and
 \begin{equation}
 g(R,t)=0\,,
 \end{equation}
 \begin{equation}
 \Lambda(R,t)=0\,\,\rm{if}\,\,
 R>R_{c}\,.\label{eq:infallangular}
 \end{equation}
where $\dot{M_{in}}$ = (m$_{0}$c$_{sc}^{3}$/G) is the infall rate and
R$_{c}$ is the centrifugal radius defined in equation
\ref{eq:centri}. Inserting equations
\ref{eq:infallmass}-\ref{eq:infallangular} into equations
\ref{eq:1}-\ref{eq:3}, we can simulate the disk evolution under the
infall.

\section{Layered accretion with infall}
In all cases we start with M$_{c}$= 0.1 M$_{\odot}$ and a cloud
mass of 1 M$_{\odot}$. We adopt a temperature of 20K for the envelope.
This parameter enters in setting the infall rate
$\dot{M_{in}}$=$c_sc^3/G = 4.2 \times 10^{-6} \msunyr$; thus the
infall lasts 2.4 $\times 10^{5}$~yr, roughly consistent with
observations (Kenyon et al 1990, 1994; Evans 2009). This leaves only
one parameter to vary, the initial angular velocity of the cloud
$\Omega_{c}$, which affects the maximum centrifugal radius onto the
resulting disk. For a singular isothermal sphere (SIS) core,
\begin{equation}
R_{cmax}\approx\frac{G^{3}M_{co}^{3}\Omega^{2}}{16
c_{sc}^{8}}\,,\label{eq:rcmaxx}
\end{equation}
where we have assumed that the central star mass
at the end of infall is approximately the cloud core
mass (M$_{co}$); thus
\begin{equation}
R_{cmax}\approx 12\, \rm{AU} \left(\frac{\Omega_{c}}{10^{-14}\rm{rad}\,\rm{
s}^{-1}}\right)^{2}\,.\label{eq:rcmax2}
\end{equation}
Since R$_{cmax}$
directly constrains the disk size after the infall, it is the most
important parameter determining the initial disk properties.

\subsection{Fiducial model}
We take as our fiducial model one with $\Sigma_{A}$=100
g cm$^{-2}$, $\alpha_{M}$=0.01, and $\Omega_{c}$=$10^{-14}
\rm{rad}\,\rm{s}^{-1}$ so that R$_{cmax} = 12$~AU.

\begin{figure}
\epsscale{.80} \plotone{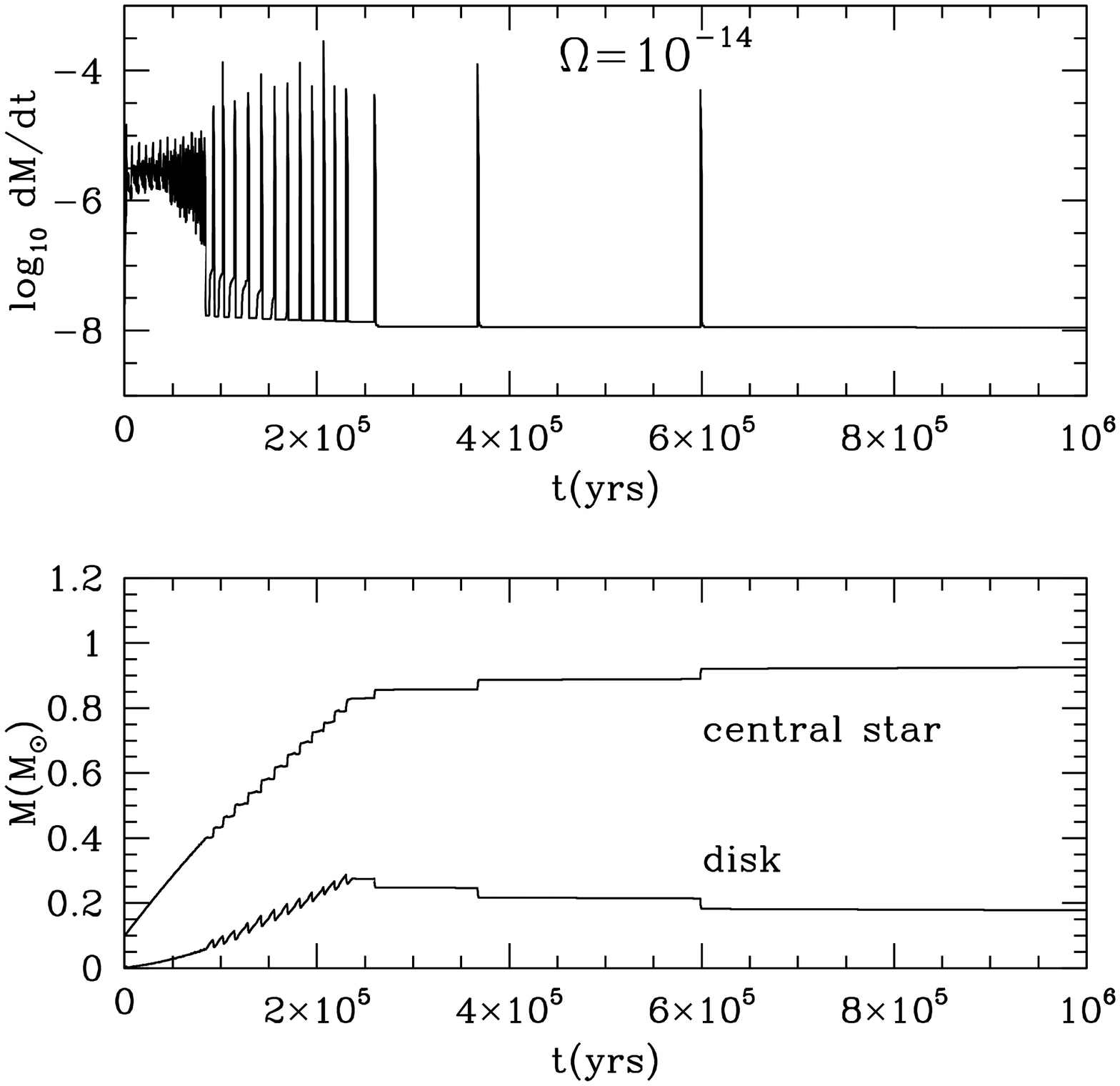} \caption{The disk mass accretion
rate (upper panel) and the mass of the central star and disk (lower
panel) with time for our fiducial model ($\Omega_{c}=10^{-14}$rad
s$^{-1}$). The upper curve in the lower panel represents the central
star mass while the lower one represents the disk mass.  }
\label{fig:dm}
\end{figure}

Figure \ref{fig:dm} shows the fiducial model's mass accretion rate
and the central star and disk masses as a function of time.  The overall
evolution can be divided into three
distinct stages. Initially there is a phase
of quasi-steady disk accretion, with the accretion rate matching the
infall rate ($\dot{M_{in}}=4.2 \times 10^{-6} \msunyr$).  This
occurs because the infall is to small radii
where the disk can become hot enough to sustain the MRI
thermally.  As infall proceeds, thermal activation of the MRI cannot be
sustained, and gravitational instability takes over as the main mechanism
of mass transfer.  However, the GI is inefficient at these radii and so
matter piles up.  Eventually, the disk becomes opaque enough that the radiative
trapping of the energy dissipated by the GI raises the central disk temperature
sufficiently to thermally-activate the MRI, and an outburst of accretion occurs.
As shown in Paper I, the result is
cycles of outbursts transferring material inward;
the GI moves material in slowly until enough mass is accumulated
with enough viscous dissipation to trigger the thermal MRI
outbursts. Finally, after infall stops, the disk is not massive
enough to sustain GI, and it accretes only from the MRI active
layer. Since the outer disk has a surface density smaller than the
maximum surface density that can be ionized by cosmic and/or X rays,
it becomes fully viscous due to the MRI; this leads to continued
expansion of the outer disk, with a slow draining of material into
the inner disk and central star. During the T Tauri phase, the
layered accretion itself may also trigger outbursts, depending on
the detailed layered structure (see below).

\begin{figure}
\epsscale{.80} \plotone{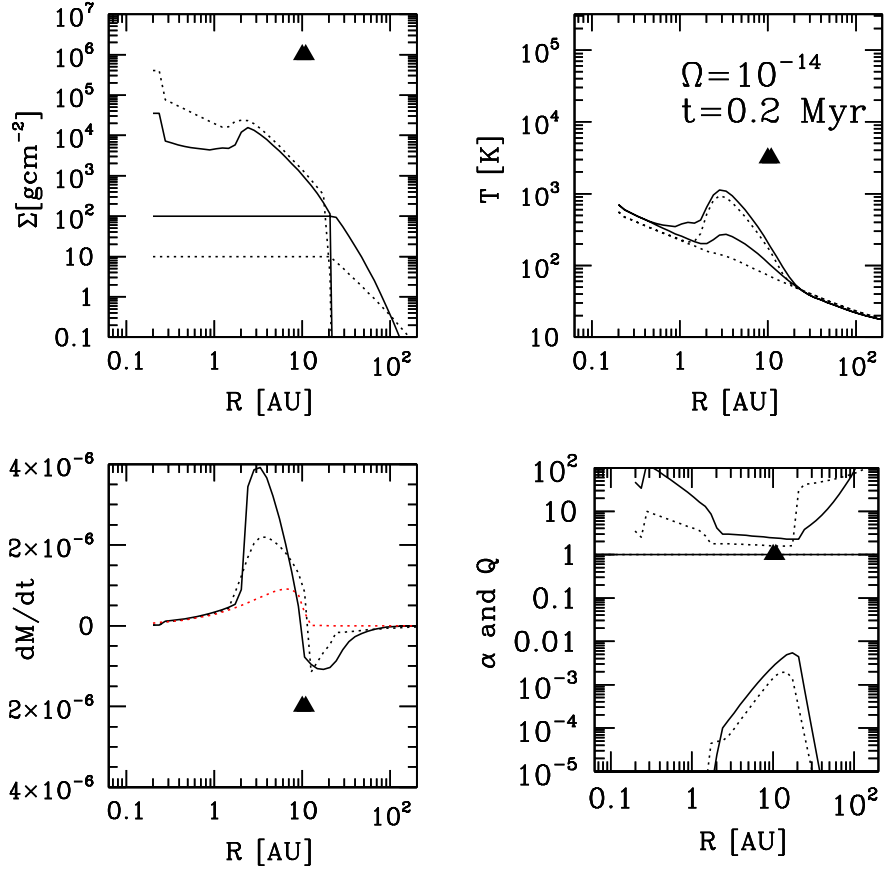} \caption{The disk surface density,
temperature structure, mass accretion rate and $\alpha$ and
Q-parameter at the end of the infall phase (0.2 Myr) for our
fiducial model ( solid curves, $\Sigma_{A}=100$g cm$^{-2}$ and
$\alpha_{M}$=0.01) and the model with a less massive active layer
but a stronger MRI viscosity (dotted curves, $\Sigma_{A}=10$g
cm$^{-2}$ and $\alpha_{M}$=0.1). The triangle labels the centrifugal
radius at this time. In the upper left panel, the curves with higher
surface densities within 20 AU represent the dead zone surface
density, while the ones with $\Sigma<$100 g cm$^{-2}$ and extend
much further represent the active layer surface density. The upper
curves in the upper right panel show the midplane temperature while
the lower curves show the active layer temperature. The black curve
in the lower left panel shows the disk mass accretion rate, while
the red curve shows the portion of the mass accretion rate lead by
the infall (the third term in equation \ref{eq:3}). The upper curve
in the lower right panel shows the Q parameter while the lower curve
shows the midplane $\alpha$ (caused by GI).} \label{fig:170}
\end{figure}

\begin{figure}
\epsscale{.80} \plotone{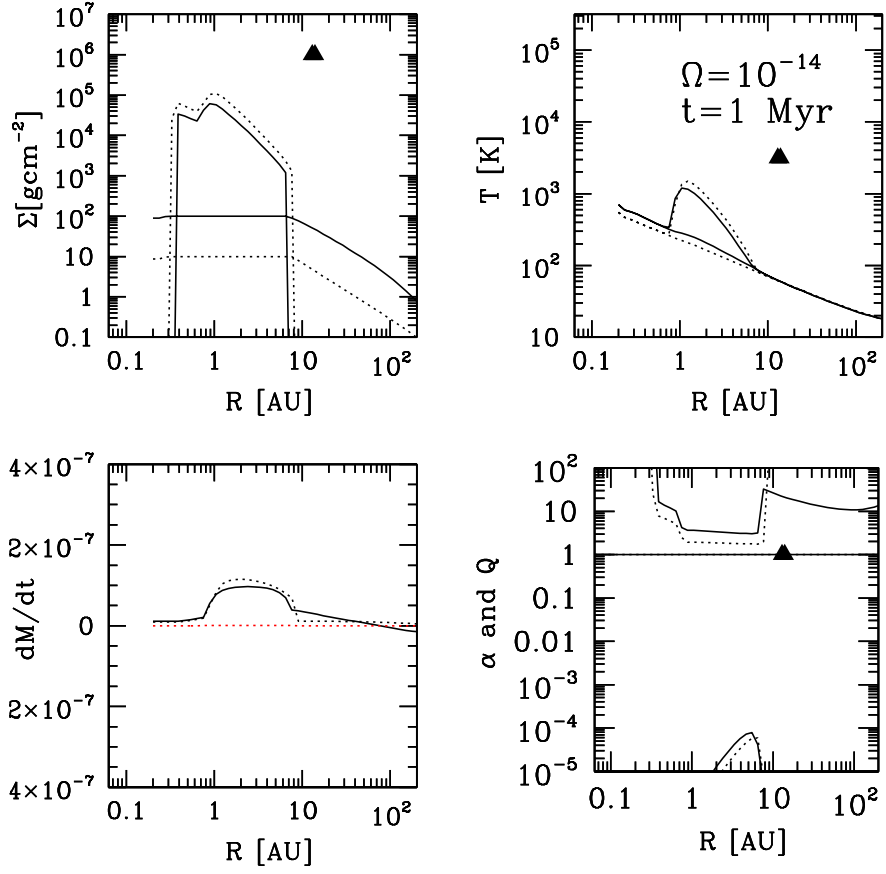} \caption{The same as
figure\ref{fig:170} but at t=1 Myr. The triangle labels the maximum
centrifugal radius.} \label{fig:300}
\end{figure}

The disk structure during the outburst stage is illustrated in
Figure \ref{fig:170}, shown at the time when the centrifugal radius
(R$_{c}$) is 10 AU (labeled by the triangle). Although the infall
ends at $R_{c}$, the disk extends far beyond $R_{c}$ due to the
outward mass transfer by the active layer. With the continuous
infall, the disk is marginally gravitationally stable within
R$_{c}$, and the energy dissipation due to the GI has significantly
heated the disk so that the outburst will be triggered at 2 AU (the
upper right panel shows that T$_{c}$ is approaching 1500 K). At the
time when infall ends, the region between $\sim$1 AU and $\sim$20 AU
is still marginally gravitationally unstable, although the disk
within 1 AU has been depleted by the previous outburst (Figure
\ref{fig:300}).

The innermost disk (R$<$0.1 AU) is purely MRI active due to
the high temperature there produced by viscous heating and irradiation.
Due to the efficient mass transport by the MRI, this inner region
can be depleted rapidly, but is limited by the accretion from
the outer active layer. Eventually, a
balance is achieved, and the inner disk behaves like a
constant-$\alpha$ disk with mass accretion rate equal to the mass
accretion rate of the innermost active layer (e.g., Gammie 1996).

In the standard layered accretion model, the disk mass accretion
rate is set as the active layer accretion rate at the dead zone
inner radius where the MRI becomes thermally activated (e.g., Gammie
1996). We use the modification of the accretion rate including
irradiation derived by Hartmann et al. (2006; see also the Appendix)
\begin{eqnarray}
\dot{M}&=&6.9\times10^{-9}\left(\frac{R}{0.2
AU}\right)\left(\frac{M_{c}}{1
M_{\odot}}\right)^{-1/2}\left(\frac{\alpha_{M}}{10^{-2}}\right)\times\nonumber\\
&&\left(\frac{\Sigma_{A}}{100 \rm{g}\,
\rm{cm}^{-2}}\right)\left(\frac{L_{*}}{L_{\odot}}\right)^{1/4}\left(\frac{f}{0.1}\right)^{1/4}
M_{\odot} yr^{-1}\,,\label{eq:layer2}
\end{eqnarray}
where $f$ is defined in equation \ref{eq:text}. At the sublimation
radius, f may be significantly larger than 0.1. With our numerical
inner boundary 0.2 AU, the simulation shows the disk has a mass
accretion rate 10$^{-8}\msunyr$ in the quiescent state, which agrees
with the above estimate.

As discussed in Paper I, we find unstable behavior at the dead zone
inner radius because of the rapid change in $\alpha_M$. This
accounts for the modest but significant variability seen during the
quasi-steady state\footnote{Although we avoided this instability at
the T Tauri phase by choosing the inner numerical boundary at 0.2
AU, this instability is present in our simulation in the first
10$^{5}$ years (short variations in Figure \ref{fig:dm}) when the
 dead zone inner radius is larger than 0.2 AU. }. This instability cannot be
reliably treated in our model, due to the complexity of the
processes involved including MRI activation and dust sublimation;
thus the precise nature of the variations in accretion during this
phase is uncertain.

A massive, marginally gravitationally unstable dead zone remains well after
the end of infall (upper left and lower right panels in Figure \ref{fig:300}).
This dead zone is maintained because the active layer mass accretion rate
increases with radius (Gammie 1996).  From the mass conservation
equation one can derive the time dependence of the total surface
density $\Sigma$=$\Sigma_{A}$+$\Sigma_{d}$,
\begin{eqnarray}
\frac{\partial \Sigma}{\partial t}&=&-\frac{\partial}{R\partial R}\left(\frac{\dot{M}}{2\pi}\right)\nonumber\\
&=&0.05 \left(\frac{1 \rm{AU}}{R}\right)\left(\frac{M}{1
M_{\odot}}\right)^{-1/2}\left(\frac{\alpha_{M}}{10^{-2}}\right)\left(\frac{\Sigma_{A}}{100
\rm{g}\, \rm{cm}^{-2}}\right)\times\nonumber\\
&&\left(\frac{L_{*}}{L_{\odot}}\right)^{1/4}\left(\frac{f}{0.1}\right)^{1/4}
\rm{g}\,\rm{cm}^{-2} \rm{yr}^{-1}\,.\label{eq:layer3}
\end{eqnarray}
Thus $\partial \Sigma\ / \partial t >0 $ and $\propto R^{-1}$. The
disk surface density thus increases linearly with time, more rapidly
at smaller radii. When the surface density
increases to the extent that the disk becomes gravitationally
unstable, self-gravity transfers the excess mass and
self-regulates the disk to an approximate Q$\sim$1 state. From equation
\ref{eq:layer3} one can see that it only takes 10$^{5}$ yr for the disk at 10 AU
to become gravitationally unstable by layered accretion.  Thus,
active-layer accretion can maintain a dead zone that is marginally
gravitationally unstable well after infall stops.

A sharp density drop appears at the outer radius of the dead zone
R$_{D}$ (also shown in Reyes-Ruiz(2007) and Matsumura et al.(2009))
and R$_{D}$ starts at the maximum centrifugal radius after the
infall and gradually moves inward with time, so that the dead zone
becomes increasingly narrow as the system evolves (see the
Appendix). In the outer disk beyond R$_{D}$ where only the active
layer is present, the disk surface density evolves from a
$\Sigma\propto R^{-1}$ distribution to an asymptotic $\Sigma\propto
R^{-1.5}$ distribution at large times. This result can be understood
by modeling it as a pure $\alpha$ disk but with a fixed $\Sigma$
inner boundary, as discussed in the Appendix.

\subsection{Varying core rotation}
The initial disk size is constrained by the maximum infall radius
R$_{cmax}$.  We next consider the results for
$\Omega_{c} = 3\times10^{-15}$rad s$^{-1}$ and $2\times10^{-14}$rad s$^{-1}$.

\begin{figure}
\epsscale{.80} \plotone{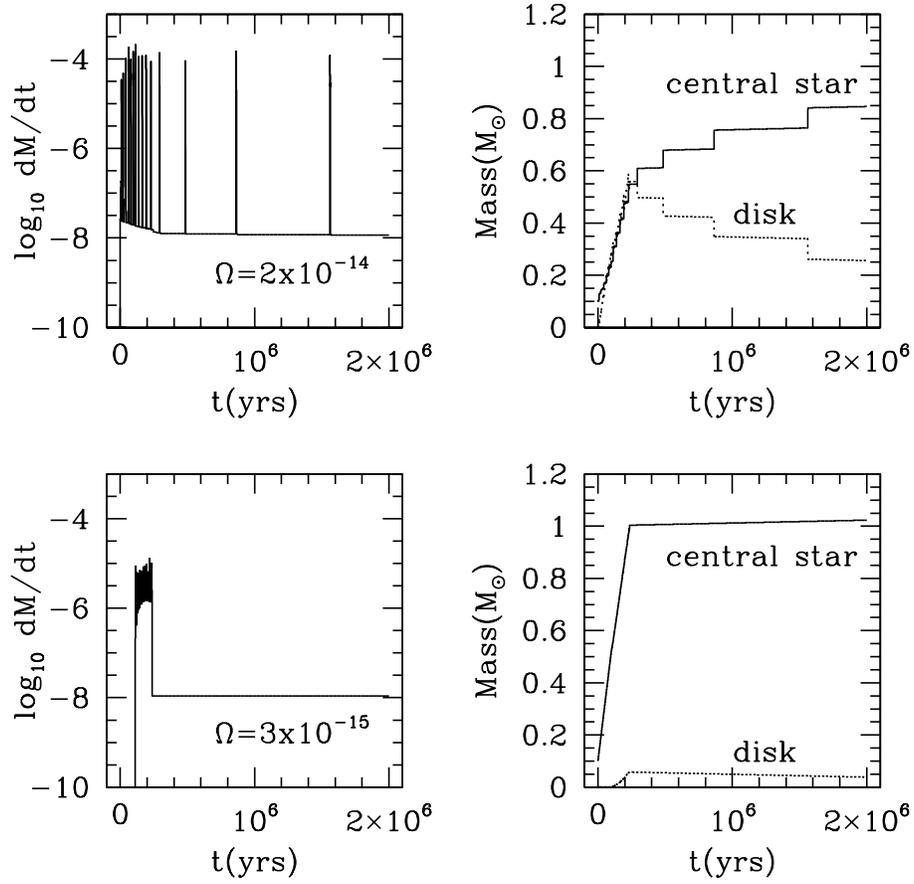} \caption{The disk mass accretion
rates and the mass of the central star and disk with time for
different core rotations. The upper two panels show the simulation
with $\Omega_{c}=2\times10^{-14}$rad s$^{-1}$, while the lower ones
with $\Omega_{c}=3\times10^{-15}$rad s$^{-1}$. The solid curves in
the right panels show the mass of the central star with time while
the dotted curves show the mass of the disk (dotted curve in the
lower right panel is too close to 0 and hard to be seen).  }
\label{fig:highlowomega}
\end{figure}

Figure \ref{fig:highlowomega} shows the accretion behavior of these
models during the first 2 Myr. In the slowly-rotating case,
all of the mass is accreted in a quasi-steady fashion, and the remaining
disk mass is very small, whereas the collapse of the rapidly-rotating
core results in very little quasi-steady accretion, an extended
period of outbursts lasting well beyond the end of infall, and a
very massive disk.

The differing behavior of the three models can be understood with
reference to Figure \ref{fig:m1}, which is a modification of the
$\dot{M}$ vs. $R$ plane discussed by Zhu \etal (2009b).  For the
slowest-rotating core model, R$_{cmax} \sim 1$~AU (equation
\ref{eq:rcmax2}). Reference to Figure \ref{fig:m1} shows that this
radius is just within the maximum radius $R_M$ for steady MRI-driven
accretion at $\dot{M} = 4 \times 10^{-6} \msunyr$.  In this region,
the disk can maintain a temperature greater than the critical value
for MRI activation; the disk then accretes on to the star as a
typical $\alpha$-disk, with modest oscillations due to the
instability at the inner disk edge described in the previous section
(see also Paper I). In the case of our fiducial model, R$_{cmax}
\sim 12$~AU; this places the disk well beyond the region where the
mismatch between GI and MRI occurs, resulting in outbursts.  The
fiducial model accretes quasi-steadily for the first $8 \times
10^4$~yr; at this point $R_c \sim 1$~AU.  For the rapidly-rotating
case, R$_{cmax} \sim 50$~AU; very little mass infalls to radii
within the MRI-steady region, and most of the mass then gets
accreted during outbursts. Because the mass transport is dominated
by the GI at large radii, a relatively massive disk results.  Note
that the outer regions of the disk reside near the limit where
gravitational fragmentation rather than simple transport might occur
(Appendix; \S 4.4).

Thus the range of rotations chosen here (given our assumed model for
the initial cloud core) approximately span the extremes of disk
evolution; at smaller $\Omega_{C}$, the accretion is almost
spherical and occurs quasi-steadily, while for larger values of
$\Omega_c$ gravitational fragmentation is likely to occur,
considerably modifying disk evolution.

The disk mass and surface density distribution are also
significantly affected by the initial core rotation. With
$\Omega_{c}=2\times10^{-14}$rad s$^{-1}$ (left panel in Figure
\ref{fig:difsurf}), although the dead zone shrinks from 35 AU at the
end of the infall phase (0.3 Myr) to 10 AU at the T Tauri phase (1
Myr), the disk is very massive (0.3 M$_{\odot}$) and still
gravitationally unstable within 10 AU during the T Tauri phase.
However, for smaller rotation $\Omega_{c}=3\times10^{-15}$rad
s$^{-1}$ (right panel), the dead zone is significantly smaller (2
AU) and has less mass (0.05 M$_{\odot}$).

When the initial core rotates slowly and R$_{cmax}$ is small, the
dead zone size is determined by R$_{cmax}$. However, if the inital
core rotates fast and R$_{cmax}$ is large, the dead zone size is
constrained by the radius where the active layer becomes
gravitationally unstable, as discussed further in the Appendix.

One important point is that even though the dead zone is small $<
50$AU, it can contain a significant fraction (or most) of the disk
mass. Figure \ref{fig:difmass} shows that, no matter what the infall
it is, most of the disk mass can be at small radii, rather than the
large radii probed by current mm and sub-mm observations (in
comparison with constant-$\alpha$ models).

\subsection{Different disk configurations}

The ionization structure of disk surface layers depends on the dust
size spectrum and the flux of ionizing radiation and cosmic rays,
and is therefore poorly constrained.  Angular momentum transport in
magnetically decoupled disks is also poorly understood, although
recent work by Lesur \& Papaloizou(2009), following Petersen et
al.(2007), strongly suggests that a nonlinear instability driven by
baroclinicity and radiative diffusion may give rise to hydrodynamic
turbulence and angular momentum transport. Therefore we have
investigated disks with a reduced active layer (\S 3.3.1), and a
residual dead zone viscosity (\S 3.3.2).

\subsubsection{Disks with less active layer and pure GI disks}

Little is known about the disk active layer, which is dependent not only
upon cosmic and/or X ray fluxes, but on the presence of dust which can
absorb charges and reduce the MRI or even prevent it (Sano \etal 2000).
Ambipolar diffusion at the low density active layer beyond 10 AU may
further depress the effectiveness of the MRI (Chiang \& Murray-Clay 2007).
In addition, there may be no dynamo in the outer disk. Thus we consider
disks with lower active-layer surface densities ($\Sigma_{A}<100$g
cm$^{-2}$).

The long-term  disk evolution is determined by the product of
$\alpha_{M}\Sigma_{A}$ (equations \ref{eq:layer2} and
\ref{eq:layer3}); thus, as long as the product
$\Sigma_{A}\alpha_{M}$ stays the same, disk evolution is similar.
Figures \ref{fig:170} and \ref{fig:300} show disks with
$\Sigma_{A}=100$ g cm$^{-2}$ and $\alpha_{M}=0.01$ and disks with
$\Sigma_{A}=10$ g cm$^{-2}$ and $\alpha_{M}=0.1$ have similar
surface density, temperature, and mass accretion rate distributions
at all times, only except that a lower $\Sigma_{A}$ leads to a less
massive outer disk.

If $\alpha_{M}\Sigma_{A}$ is lower, the disk accretion rate is lower
than our fiducial model; thus the infall mass takes longer to
accrete onto the star. In addition, with a smaller $\Sigma_{A}$, the
dead zone outer radius (R$_{D}$) is larger (equation \ref{eq:rt}).

In the extreme case when there is no active layer \footnote{We allow
 thermal activation of the MRI when the disk temperature is higher
than T$_{M}$.}, the disk evolves towards Q$\sim$1 during the
simulation (Figure \ref{fig:deadzonevis}), since the GI is the only
mechanism to transport mass from the outer disk inwards. The disk is
very massive and has a sharp outer disk edge roughly at R$_{cmax}$
(Vorobyov \& Basu 2005, 2006, 2007). Without the active layer
replenishing the innermost disk and the inner disk becoming
gravitationally stable (due to the strong irradiation from the
central star), the disk accretion rate drops to zero (upper left
panel in Figure \ref{fig:deadzonevis}). During the T Tauri phase,
the disk evolves to Q$\sim$2 and stays at this gravitationally
stable state (Figure \ref{fig:deadzonevis}). Thus the disk is
massive (0.2 M$_{\odot}$) during the T Tauri phase. Even if the disk
has some small residual viscosity at Q$>>$1, it has limited effect
on disk evolution simply because, with $\alpha<10^{-4}$, the viscous
timescale is too long (more than 2 Myr to transport mass from 50 AU
to the central star) (see also Vorobyov \& Basu 2009).

\subsubsection{Residual dead zone viscosity}

In our fiducial model, we assume the dead zone has no viscosity at
all. However, numerical MHD simulations have shown that the dead
zone may not be totally ``dead'', with a small residual viscosity.
The viscosity could be due to the operation of the MRI with the ions
turbulent mixing from the active layer (Ilgner \& Nelson 2008) or
from turbulence penetrating the dead zone from the active layer
(Turner \etal 2007). Recent work by Lesur \& Papaloizou (2009) also
persuasively demonstrates the existence of a nonlinear instability
driven by baroclinicity and radiative transport (the magnitude is
not yet clear, and the existing evidence suggests that the
instability weakens when radiative diffusion is small---when the
Peclet number is large---as might be expected in a dead zone with
large optical depth.

If the residual viscosity in the dead zone is $\alpha_{rd} \sim$
10$^{-5}$), it has a negligible effect on disk evolution.  Disk structure
evolves simply on the viscous timescale; for $\alpha=10^{-5}$, this is
long compared to T Tauri lifetimes $\sim$ few $\times$10$^{6}$ yr.

However, if the residual $\alpha_{rd}\gtrsim$10$^{-4}$, the dead zone can
transport the mass at a rate comparable to the accretion rate of the
active layer, with $\dot{M_{i}}\propto\Sigma_{i}\alpha_{i}$ and
$\Sigma_{d}>>\Sigma_{a}$. We assume the disk has a constant mass accretion
rate after a long period of time, and at the stellar surface
$d\Omega/dR$=0. Furthermore, if the disk is irradiation-dominated such
that T$\propto$R$^{-1/2}$, $\nu_{a}$=k$_{a}$R and $\nu_{d}$=k$_{d}$R where
k$_{a}$ and k$_{d}$ are constants and are proportional to $\alpha_{a}$ and
$\alpha_{d}$, we derive (in the Appendix)
\begin{equation}
\Sigma_{d}=\frac{\dot{M}}{3\pi k_{d} R}
\left[1-\left(\frac{R_{*}}{R}\right)^{1/2}\right]-\frac{k_{a}}{k_{d}}\Sigma_{a}\,.\label{eq:sigmad}
\end{equation}
If the dead zone transports mass at a rate higher than the active
layer, the last term can be neglected for an order of magnitude
argument. Compared with a constant-$\alpha$ disk having the same
mass accretion rate, the layered disk has a surface density higher
than the constant-$\alpha$ disk by a factor of
$\alpha$/$\alpha_{d}$. If we assume the outer pure MRI active disk
and the inner layered disk has the same mass accretion rate, and the
outer disk only has $\nu_{a}$ while the inner dead zone is dominated
by $\nu_{d}$, the disk's surface density will change by a factor of
$\alpha_{a}$/$\alpha_{d}$ within the dead zone outer radius R$_{D}$,
which is shown in Figure \ref{fig:deadzonevis2}. Also this density
change at R$_{D}$ is more gradual (equation \ref{eq:sigmad} is a
smooth function of R) than the zero dead zone viscosity case. This
can have a significant impact on planet migration (\S 4.6).

\section{Discussion}
\subsection{Dead zones and star formation by disk accretion}

The main result of our work is that star formation through disk
accretion, assuming a sufficiently large initial disk that the MRI
is not thermally activated throughout the disk, generally results in
a dense belt of material - the ``dead zone'' -which has implications
for disk structure and planet formation.

To demonstrate this more clearly, we show the results for a
constant-$\alpha$ disk for the same infall model. By comparison with
layered disks in Figure \ref{fig:difmass}, constant-$\alpha$ disks
have similar outer disk but lack the massive dead zone.
Constant-$\alpha$ disks respond to infall just as layered disks do:
a faster rotating core leads to a more massive disk since the infall
mass onto larger radii takes longer to accrete to the central star.
Our models show that with the same $\alpha$ in constant-$\alpha$
models, cores with $\Omega_{c}=2\times10^{-14}$rad s$^{-1}$ lead to
disks 10 times more massive than the disks produced by cores with
$\Omega_{c}=3\times10^{-15}$rad s$^{-1}$ at 1 Myr.  However, even if
we allow constant-alpha disks to have different values of alpha
(Figure \ref{fig:constalphadm}), all these models produce surface
densities far in excess of 100 g cm$^{-2}$ over large areas of the
disk at the end of infall (lower left panel of Figure
\ref{fig:constalphadm}). Thus, {\em a reduced angular momentum
transport efficiency associated with the failure of the MRI at
$\Sigma > 100$g cm${-2}$ implies the formation of a dense belt or
dead zone in the disk}.

Another feature of the constant-$\alpha$ model is that, although the
core rotation determines the disk mass, the disk surface density
shape depends most strongly on $\alpha$. This can be simply
explained by the similarity solution of the constant-$\alpha$ disk
where the shape of the disk is determined by the scaling radius
(Hartmann et al. 1998), which is proportional to t and $\alpha$.
Thus as long as t and $\alpha$ are the same, the disk surface
density has similar shape. We also notice that, even for a
constant-$\alpha$ disk, the disk is gravitationally unstable at the
infall phase with a reasonable set of parameters ($\alpha$=0.01 and
$\Omega_{c}$=2$\times$10$^{-14}$rad s$^{-1}$) (Figure 10).

\subsection{Relation to previous work}

As mentioned in the introduction, previous investigations by
Vorobyov \& Basu (2006, 2007) and Kratter \etal (2008, 2009) have
considered the evolution of disks formed by infall from rotating
cloud cores with angular momentum transport by GI. Vorobyov \& Basu
find outbursting behavior, as we do, but for different reasons;
specifically, in their models time-dependent accretion is driven by
gravitational clumping, which we cannot treat in one dimension.
However, with more realistic energy equations, it is harder to lead
to strong gravitational clumping. Kratter \etal (2009) studied more
generalized models of infall; they did not find clumping but found
conditions for which fragmentation would occur, which agrees with
our estimate in Appendix B.

Unlike these investigations, we study slowly rotating cloud cores
forming single stars without disk fragmentation, focussing on disk
evolution after the central core collapse.  We treat the GI as a
local phenomenon, which may be problematic if the disk is very
massive (Lodato \& Rice 2005). However, during later stages when the
disk is less massive, a local treatment appears to be reasonable
(Lodato \& Rice 2004, Cossins \etal 2009a). In addition, we are able
to treat radiative cooling more realistically; the treatment of
thermal physics plays a crucial role in GI transport (e.g., Durisen
\etal 2007) and MRI-GI outbursts (Paper I).

Lodato \& Rice (2005) found recurrent episodes of gravitational
instability in a massive disk. Boley \& Durisen (2008) considered
three-dimensional simulations of gravitationally-unstable disks with
accurate radiative transfer, and also suggested that high mass
fluxes could result from the rapid onset of GI, producing something
like an FU Ori outburst.  The question is whether this behavior is
 a transient due to an initial
condition, which may or may not be realizable as part of a natural
evolutionary sequence. This GI outburst could be a boost in the MRI
activation.

Rice \etal (2009) also considered the evolution of disks formed by
infall with GI as the transport mechanism.  They also find that
massive disks result, which leads them to suggest that another
mechanism of viscosity must be operating to drain disks.  We address
this possibility in \S 4.5.

Our treatment differs from all of the above in including MRI transport
with layered accretion. This results in our outer disks tending to evolve
viscously rather than through GI, so that mass accretion can continue even
after the disk is gravitationally stable.  Perhaps most importantly, our
one-dimensional model allows us to treat the very innermost disk at the
same time considering the outer disk evolution, something that is
extremely difficult to do with two- or three-dimensional codes.  This
allows us to demonstrate behavior not seen in the other simulations;
specifically, the onset of MRI-driven accretion in the inner disk, which
tends to dominate the outburst behavior - or, as we have shown, result in
a phase of quasi-steady disk accretion while low angular momentum core
material is infalling.

At high
accretion rates, it appears almost certain that the MRI should
operate, as temperatures become high enough to sublimate dust, which
might otherwise absorb charges.  This notion is supported by
observations of FU Ori (Zhu \etal 2007, 2008), which indicate that
the disk can be hot enough to eliminate dust out to radial scales of
order 1 AU.  Thus we argue that inclusion of the MRI is important
for any understanding of accretion onto central protostars.

The behavior of non-thermally activated MRI by X-ray or cosmic ray
radiation is much more uncertain, but its inclusion can also have
important effects not seen in pure GI treatments.  In our layered
simulations, during the T Tauri phase the GI has limited effects on
the long term disk evolution and the disk mass accretion rate is
controlled by the active layer, as discussed in \S 3. To confirm
this, we have run the same simulation as before but with the
$\alpha_{Q}$ from Armitage et al. (2001), which is based on Lin\&
Pringle (1987, 1990). We found that the different forms of
$\alpha_{Q}$ have little effect on the long term evolution of a
layered disk; the T Tauri disk evolution is more determined by the
active layer. In both treatment of $\alpha_{Q}$, the layered
accretion leads to a relatively massive ``inner belt'' within 30 AU
and a viscous disk beyond 30 AU (Figure \ref{fig:300}).

\subsection{Outbursts and luminosity problem}

Our model predicts that protostars will undergo FU Ori-like
outbursts of rapid mass accretion to accumulate a significant amount
of their mass, helping to solve the luminosity problem (\S 1).  Here
we look at the question of luminosity evolution more closely.

The largest current survey of star-forming regions bearing upon this
issue is that of the c2d Spitzer Legacy project (e.g, Enoch \etal
2009, Evans \etal 2009), which strongly suggests that protostellar
accretion is time variable, with prolonged periods of low accretion
rates. Evans \etal (2009) estimate that half of the mass of a
typical low-mass star is accreted during only $\sim$~7\% of the
Class I lifetime, which they estimate as $\sim 0.5$~Myr. In our
fiducial model, roughly half of the mass is accreted quasi-steadily
during the first $\sim 10^5$~yr, with the other half being accreted
during outbursts within the next $\sim 1.5 \times 10^5$~yr. This
agrees with the c2d observations that half of the mass is accreted
during high mass accretion rate stage. However, the outburst
behavior (episodic accretion) for the individual protostar does
depend on the initial rotation of the cloud core. The $\Omega_c = 2
\times 10^{-14}$ case accretes almost entirely via outbursts, while
the low-rotation case accretes essentially all of the stellar mass
quasi-steadily.

Evans \etal (2009) found that 59\% of Class 0/I stars have
bolometric luminosities smaller than 1.6 L$_{\odot}$. However, in
our fiducial model, 30\% of the Class 0/I phase is in the
quasi-steady accretion phase with luminosity $\sim$13 L$_{\odot}$
(GM$_{*}\dot{M}$/R$_{*}$ by assuming
$\dot{M}$=4$\times$10$^{-6}\msunyr$ and
M$_{*}$/R$_{*}$=M$_{\odot}$/10R$_{\odot}$), much larger than 1.6
L$_{\odot}$. One possible solution of this problem is that the
extinction toward the protostar during this quasi-steady phase is
high, with A$_{V} \gtrsim$ 400 using our model parameters; it might
thus be difficult to identify these objects in such early stages of
evolution.

 Another possible solution is the oscillation of the
thermal MRI region (W{\"u}nsch et al. 2006) can lead to additional
variations in the mass accretion rate and accretion luminosity
(which is the reason we call the first steady accretion phase
``quasi-steady''; variations are clearly seen in Figure \ref{fig:dm}
during the first 10$^{5}$ years)(see \S 3.1). We have avoided
concentrating on this feature because it depends on the complex
physics at the inner edge of the dead zone, which is unclear to us
now. Nevertheless, this potentially unstable behavior may be
important in understanding protostellar accretion in general and may
even provide a mechanism for explaining EXor outbursts (Herbig
1977). Further investigation of this problem with a realistic
calculation of the onset of the MRI is needed.

Some other limitations of our models are that the infall is
axisymmetric with a constant infall rate, and that the GI is treated
as a local effect. If the infall has a complex pattern and the GI
has a strongly global character, the angular momentum transport
during the infall phase could be very efficient (Larson 1984) and
disk accretion may be more variable (Vorobyov 2009) or even form
warped disk (Bate et al. 2009).

There is also a distinction between the model predictions
for outbursts during the infall phase and those
which occur after infall ends, during the T Tauri phase. Though they
are both triggered by mass accumulation, this is
caused by GI during the infall phase, which is independent of the layered
structure (Paper I), while the outbursts during the T Tauri phase are
triggered by the mass accumulation from active layer
accretion, which is sensitive to the assumed value of $\Sigma_a$.

\subsection{Rotation and Fragmentation}

Our most rapidly-rotating model formally predicts FU Ori-like
outbursts during the T Tauri phase, driven by active layer
accretion. However, this T Tauri phase outburst behavior is
uncertain because it sensitively depends on active layer properties
and the massive disks which would produce such behavior fragment
instead into multiple stellar systems with individually smaller
disks (Rapidly-rotating cores tend to produce large disks that are
subject to gravitational fragmentation at R$ \gtrsim$ 50 AU in a
realistic disk, see Rafikov 2009; Cossins et al.  2009b; see
Appendix).

Finally, we consider possible values of envelope angular momentum.
Typical observational estimates of the ratio of rotational kinetic to
gravitational energy span a range of three orders of magnitude (e.g.,
Caselli \etal 2002), and thus are difficult to apply to our simulations.
Moreover, the total angular momentum is extremely sensitive to assumptions
about the distribution of mass with radius as well as the (not necessarily
uniform) angular velocity; as cores are generally not Bonnor-Ebert
spheres, or even axisymmetric about the rotation axis (Tobin \etal 2009),
making it difficult to compare with observations.  As mentioned above,
cores with more angular momentum than our fastest-rotating model are
likely to fragment into multiple systems.  Whether significant numbers of
slowly-rotating cores exist which can produce diskless T Tauri stars, or T
Tauri systems with very low-mass disks, will require more detailed
observational analyses of the complex geometries of starless cores.

\subsection{Dead zones: do they exist?}

We have shown that a natural consequence of our layered model is the
formation of dead zones that may remain relatively massive for more than
$10^6$ yr (also Vorobyov \& Basu 2006, 2007; Rice \etal 2009).  This
stands in contrast with purely viscous models of disk evolution, as shown
in Figure \ref{fig:difmass}.  An advantage of highly-viscous models is
that they provide an explanation for disk accretion rates during the T
Tauri phase (Hartmann \etal 1998; Dullemond, Natta, \& Testi 2006),
whereas the layered accretion model has some difficulties in explaining T
Tauri accretion rates (Hartmann \etal 2006).  On the other hand, the pure
viscous models are unable to explain FU Ori outbursts, which require large
mass reservoirs at distance scales on the order of an AU (Armitage, Livio,
\& Pringle 2001; Zhu \etal 2007, 2008, 2009).

At present the best that can be said is that the MRI is likely to operate
at $r \lesssim 0.1$AU, and it is unlikely to operate at larger radii where
the temperature is lower.  Even the standard form of the minimum-mass
solar nebula (Weidenschilling 1977), let alone the recent amplification
suggested by Desch (2007; Figure \ref{fig:desch}), indicates surface
densities well in excess of those thought to be penetrated by cosmic rays,
or even stellar X-rays.  At any rate, the associated change (presumably
decrement) in angular momentum transport efficiency where the MRI shuts
off is likely to give rise to a feature in the surface density.

At first sight, the surface densities of the dead zones in our
evolutionary models seem unreasonably large in comparison with
previous models. However, Terquem (2008) has shown that, even for a
constant mass accretion rate layered model, the dead zone is very
massive. Reyes-Ruiz (2007) has also shown that a massive disk still
exists even if the dead zone has some residual viscosity. Some of
our models do not predict values of $\Sigma$ much higher than that
of the Desch (2007) model (Fig. \ref{fig:desch}). If the dead zone
has some residual viscosity with $\alpha>10^{-4}$, the dead zone
mass can be significantly reduced but still relatively massive
compared with a constant-$\alpha$ disk (Figure \ref{fig:submm}).
Furthermore, some exoplanetary systems have masses substantially
larger than that of Jupiter, suggesting that more than minimum mass
solar nebulae are required, especially if migratory loss of solid
bodies is significant.

Have high surface density regions been ruled out by recent interferometric
observations of submm dust emission in disks, such as those of Andrews
\etal (2009)?  To examine this possibility, in Figure \ref{fig:submm} we
compare various models (our fiducial model, fiducial model with a residual
dead zone viscosity, and a constant-$\alpha$ accretion model) with the
suite of disk structures inferred by Andrews \etal (2009). The agreement
between the Andrews \etal results and our models in the outer disk is not
surprising, because they adopted viscous disk solutions like ours to model
the observed interferometric visibilities. If we adopt the same opacity as
Andrews \etal, we find that there is no strong signal in the disk surface
brightness distribution at the edge of the dead zone because it is
optically thick and the temperature distribution is dominated by
irradiation from the central star (the brightness temperatures for the
layered model and the constant-$\alpha$ model are the same at 1 mm in
Figure \ref{fig:sb}). The model does predict a possible feature at
$\lambda \sim 1$~cm where the outer disk becomes optically thin (Figure
\ref{fig:sb}); this might be testable with the EVLA.  The high surface
densities in the dead zone may lead to rapid coagulation of the solid
particles, however, reducing the optical depth contrast between the dead
zone and the MRI-active outer disk region.

Our results suggest the potential importance of dead zones for
simply providing a large mass reservoir for solid body formation.
Higher density regions may be problematic in that timescales for
radial drift of small bodies and Type I migration become much
shorter; on the other hand, this may be offset by having more
material to start with, and other features of the dead zone may help
provide ``traps'' where planetesimal can form efficiently (Rice et
al. 2004) and migration can be halted or even reversed (see the
following section).

In this connection it is worth noting the existence of the so-called
``transition'' disks, where outer, optically-thick disks surrounding
T Tauri stars have very large inner disk holes - either partially or
fully evacuated (e.g, Espaillat \etal 2007).  In some cases these
large inner disk holes may be the result of tidal torques from
companion (binary) stars, but in others giant planets may be the
cause of the inner disk clearing.  In the latter case, it may be
necessary to have multiple giant planets form nearly simultaneously
over a range of radii in order to explain the sizes of these disk
holes. The dead zones of our models, with relatively large surface
densities over a significant range of radii, may be able to promote
the necessary runaway growth.

\subsection{R$_{D}$ and planet migration}

The sharp density jump at the dead zone outer radius (R$_{D}$) can
significantly affect the planet migration outside the dead zone (Matsumura
\& Pudritz 2007);  the associated torque on the planet depends on R$_{D}$,
the density jump factor $F$, and the width of the density jump.

In the framework of this paper, we derive the first two parameters
analytically and test it numerically. For R$_{D}$, \S 3.1 and the Appendix
have shown that it starts at the outer disk at the end of the infall, and
then travels inwards with time at a speed given in equation \ref{eq:vd} in
the Appendix.  With the initial R$_{D}$$\sim$R$_{cmax}$ (if we only
consider the slow rotating core: no GI fragmentation and active layer GI
as discussed in the Appendix) and the speed of R$_{D}$, the position of
R$_{D}$ can be derived at any time later (Figure \ref{fig:tran}). Based on
equation \ref{eq:vd}, the speed of R$_{D}$ gets slower when moving
inwards, which is seen in Figure \ref{fig:tran}.

We can derive $F$ if the dead zone residual $\alpha$ is small enough
($\alpha<10^{-5}$). With such a small dead zone residual viscosity,
the disk is gravitationally unstable within R$_{D}$. Thus the
surface density jump can be simply derived by dividing the surface
density of the Q=1 disk to $\Sigma_{A}$\,,
\begin{equation}
F=\frac{\Sigma_{Q=1}}{\Sigma_{A}}=\frac{c_{s}\Omega}{\pi
G\Sigma_{A}}=\left(\frac{45 AU}{R_{D}}\right)^{7/4}\left(\frac{100 g
cm^{-2}}{\Sigma_{A}}\right)\label{eq:F}
\end{equation}
Here we have assumed the temperature profile as $T$=200 K(R/1
AU)$^{-1/2}$. Thus, as the dead zone moves inwards ($R_{D}$ becomes
smaller), $F$ increases.  However, the density jump width, which is the
last parameter required to calculate the torque on the planet, cannot be
constrained by our 1D simulation.

\subsection{MMSN and planet formation in the layered disk}

Although at the early stage the inner disk is massive, at later
stage ($\sim$Myr) the outer disk (beyond 10 AU) is comparable to the
minimum mass solar nebulae (MMSN) from Weidenschilling (1977)
(Figure \ref{fig:desch}). Due to the boundary effect at the dead
zone outer radius R$_{D}$ (\S 3.1 and Appendix), the outer disk
evolves towards $\Sigma\propto R^{-1.5}$ as in the standard MMSN.
Furthermore, if the dead zone is massive, R$_{D}$ moves inwards very
slowly (\S 4.5); then $\Sigma\propto R^{-1.5}$ lasts for a long
time.

If planets form in a massive dead zone, they may be lost by inward
migration; however, some may be trapped at the inner boundary
(Kretke \etal 2009) or outer boundary (Matsumura \etal 2007,2009) of
the dead zone.

\section{Conclusions}

In this paper, we have constructed a one-dimensional two-zone accretion
disk model to study disk formation and long-term evolution under the
collapse of a BE rotating core.  The model evolution can be divided into
three stages. At the early stage, when the mass falls to the inner disk
within AU scale, the MRI can be sustained in the inner disk and
efficiently and steadily transfers the infalling mass to the central star.
Later, when the mass falls beyond AU scale, the disk goes to the outburst
stage due to the accretion rates' mismatch by the MRI and GI as described
in Paper I. After the infall completes, the disk enters the T Tauri phase
and evolves on its own.  Cores with higher initial rotation end up with a
more massive disk and more disk episodic accretion events (outbursts). As
long as the initial cloud core does not rotate extremely slowly to form a
tiny disk (R$_{cmax}$$\sim$1 AU), more than half of the star mass is built
up by outbursts, which eases the ``luminosity'' problem.

Disks exhibit a variety of behavior during the T Tauri phase. For a disk
with accretion sustained only by gravitational instability, the disk
evolves towards a Q=1  disk and the disk truncates at a radius slightly
larger than the maximum centrifugal radius of the infall.  If the disk has
an active layer at the surface, however, the active layer can extend to a
much larger radius and a sharp density drop develops at a characteristic
radius R$_{D}$ that separates the marginally gravitationally stable dead
zone and the MRI active but gravitationally stable outer disk.  The
density jump at R$_{D}$ may be observable by the EVLA and ALMA.  The
formation of a dense belt of material is associated with the failure of
magnetically driven transport due to low ionization at intermediate radius
in the disk; the only ways to avoid this are (1) if there is a separate,
equally efficient hydrodynamic transport mechanism, or (2) if for some
reason the MRI fails in the outer disk as well, perhaps due to dynamo
failure.

\acknowledgments

We thank Sean M. Andrews and David J. Wilner for kindly allowing us
to use their figure in our paper. ZZ thanks Robin Fowler for
carefully reading the manuscript and improving the writing. We thank
the referee Giuseppe Lodato for a helpful and thorough report. This
work was supported in part by NASA grant NNX08A139G, by the
University of Michigan, by a Sony Faculty Fellowship, a Richard and
Margaret Romano Professorial Scholarship, and a University Scholar
appointment to CG.

\appendix

\section{A. Layered accretion} For a layered disk dominated
by viscous heating, Gammie (1996) showed that the central
temperature and the mass accretion rate at the radius R are
\begin{eqnarray}
T_{c}&=&290 \left(\frac{R}{1
\rm{AU}}\right)^{-3/5}\left(\frac{M_{c}}{1
M_{\odot}}\right)^{1/5}\left(\frac{\alpha_{M}}{10^{-2}}\right)^{2/5}\nonumber\\
&&\times\left(\frac{\Sigma_{A}}{200 \rm{g}\,
\rm{cm}^{-2}}\right)^{4/5}\,\rm{K}\,,
\end{eqnarray}
\begin{eqnarray}
\dot{M}&=&1.17\times10^{-7}\left(\frac{R}{1
\rm{AU}}\right)^{9/10}\left(\frac{M_{c}}{1
M_{\odot}}\right)^{-3/10}\left(\frac{\alpha_{M}}{10^{-2}}\right)^{7/5}\nonumber\\
&&\times\left(\frac{\Sigma_{A}}{200 \rm{g}\,
\rm{cm}^{-2}}\right)^{9/5} M_{\odot} yr^{-1}\,,\label{eq:layer}
\end{eqnarray}
where the Bell \& Lin (1994) opacity has been used.

In the irradiation dominated limit, the disk temperature
is the external temperature (equation
\ref{eq:text}), $\sigma T^{4}=f L_{*}/4\pi R^{2}$. In this case the disk
temperature and mass accretion rate with radii are
\begin{equation}
T_{c}=221 \left(\frac{L_{*}}{L_{\odot}}\right)^{1/4}\left(\frac{R}{1
\rm{AU}}\right)^{-1/2}\left(\frac{f}{0.1}\right)^{1/4}\, K\,,
\end{equation}
\begin{eqnarray}
\dot{M}&=&6.9\times10^{-8}\left(\frac{R}{1
\rm{AU}}\right)\left(\frac{M_{c}}{1
M_{\odot}}\right)^{-1/2}\left(\frac{\alpha_{M}}{10^{-2}}\right)\nonumber\\
&&\times\left(\frac{\Sigma_{A}}{200 \rm{g}\,
\rm{cm}^{-2}}\right)\left(\frac{L_{*}}{L_{\odot}}\right)^{1/4}\left(\frac{f}{0.1}\right)^{1/4}
M_{\odot} yr^{-1}\,.\label{eq:layer4}
\end{eqnarray}
Comparing equations \ref{eq:layer4} and \ref{eq:layer}, we see that
the disk is irradiation dominated if the active layer has surface
density ($\Sigma_{A}<10^{2}$ g cm$^{-2}$), or $\alpha_{M}$ is
smaller than $10^{-2}$, or the luminosity is significantly larger
than the stellar radiation (L$>$10$L_{\odot}$).

\subsection{A.1 Dead zone} In either of the above cases the layered
disk accretion rates increase nearly linearly with radius, which
results in piling up mass in the dead zone at small radii. Using the
mass conservation equation
\begin{equation}
R\frac{\partial \Sigma}{\partial t}=-\frac{\partial}{\partial R}
\left(\frac{\dot{M}}{2\pi}\right)\,,
\end{equation}
$\partial \Sigma\ / \partial t \propto R^{-1}$. Thus the layered
disk surface density increases linearly with time and it increases
more rapidly at smaller radii. If we assume the dead zone has zero
residual viscosity (non-zero residual viscosity has been discussed
in \S 3.3.2) and active layer viscosity $\nu_{a}$=k$R^{n}$, disk
evolution becomes:
\begin{equation}
\frac{\partial \Sigma}{\partial
t}=3n\left(n+\frac{1}{2}\right)k\Sigma_{A}R^{n-2}\,.\label{eq:evolve4}
\end{equation}
Thus
\begin{equation}
\Sigma(R,t)=3n\left(n+\frac{1}{2}\right)k\Sigma_{A}R^{n-2} t+C\,,
\end{equation}
which increases linearly with time due to the layered accretion, as shown
by Matsumura \& Pudritz (2007).

The dead zone starts at 0.1 AU and ends at the radius where the disk
surface density is smaller than $\Sigma_{A}$ (the dead zone outer
radius R$_{D}$), leaving only the MRI active layer ionized by cosmic
and/or X rays at larger radii. This pure MRI active outer disk
beyond R$_{D}$ evolves like a constant-$\alpha$ viscous disk.
However since the active layer of the inner disk accretes mass
inwards, the dead zone is gradually depleted so that R$_{D}$  moves
inwards with time.

To determine the dead zone size (R$_{D}$) during the layered disk
evolution, we need to know its initial position just after the
infall. Initially R$_{D}$ should be close to R$_{cmax}$ inside of
which the infall mass lands. However this is only true if the
initial core rotates slowly and R$_{cmax}$ is small. If the inital
core rotates rapidly and R$_{cmax}$ is large, the dead zone size is
constrained by the radius where the active layer becomes
gravitationally unstable. Generally, the dead zone cannot extend to
R$>$50AU if $\Sigma_{A}$=100 g cm$^{-2}$ due to the active layer GI.
This is because the GI can be very efficient in transporting mass
and angular momentum when $Q<1$, and the surface density with Q=1
can be considered as an upper limit that the disk surface density
cannot exceed during the evolution. For a layered disk dominated by
viscous heating, Gammie (1996) has shown that the Toomre Q parameter
at R is
\begin{equation}
Q=1800 \left(\frac{R}{1 AU}\right)^{-9/4}\left(\frac{M_{c}}{1
M_{\odot}}\right)^{3/4}\left(\frac{\alpha_{M}}{10^{-2}}\right)^{1/2}\left(\frac{\Sigma_{A}}{\Sigma}\right)\,.
\end{equation}
At the outer dead zone radius (R$_{D}$), $\Sigma=\Sigma_{A}$; with
the condition that Q$>$1,
\begin{equation}
R_{D}<28 \left(\frac{M_{c}}{1
M_{\odot}}\right)^{1/3}\left(\frac{\alpha_{M}}{10^{-2}}\right)^{2/9}
AU\,.\label{eq:rt2}
\end{equation}

For an irradiation dominated disk, we derive a similar condition:
\begin{equation}
Q=\frac{c_{s}\Omega}{\pi G (\Sigma_{a}+\Sigma_{d})}\,,
\end{equation}
assuming
\begin{equation}
T(R)=200 \rm{K} (\frac{R}{\rm{AU}})^{-1/2}\,.\label{eq:T}
\end{equation}
Then,
\begin{equation}
Q(R_{D})=\left(\frac{45 \rm{AU}}{R_{D}}\right)^{7/4}\left(\frac{100
\rm{g}\,\rm{cm}^{-2}}{\Sigma_{A}}\right)\,.
\end{equation}
Since Q(R$_{D}$)$>$1,
\begin{equation}
R_{D}<45 \left(\frac{100 \rm{g}
\,\rm{cm}^{-2}}{\Sigma_{A}}\right)^{4/7} \rm{AU} \,.\label{eq:rt}
\end{equation}
In either case, R$_{D}\lesssim$ 50 AU. To test this, we computed a
case with a $\Omega_{c}=2\times10^{-13}$rad s$^{-1}$ core. The core
mass falls into R$_{cmax}$ as large as 400 AU, but the dead zone
beyond 30 AU is quickly cleared by the GI so that R$_{D}$ is smaller
than 30 AU during the T Tauri phase.

After we know the initial R$_{D}$ position, the motion of R$_{D}$
can be derived by considering the mass conservation at R$_{D}$. For
the region extending from R$_{D}$-$\Delta R$ to R$_{D}$,
\begin{equation}
R\frac{\partial \Sigma}{\partial t}+\frac{\Delta
(R\Sigma_{a}v_{R})}{\Delta R}=0\,.
\end{equation}
Assuming the mass flux (or $v_{R}$) at R$_{D}$ is zero, which will
be justified in some cases later, and considering the time for the
dead zone within $\Delta$ R to be depleted is $\Delta t$, we got
\begin{equation}
R\frac{\Sigma_{tot}}{\Delta t}-\frac{(R\Sigma_{a}v_{R})}{\Delta
R}=0\,,
\end{equation}
thus
\begin{equation}
v_{D}=\frac{\Delta R}{\Delta
t}=v_{R}\frac{\Sigma_{a}}{\Sigma_{tot}}=\frac{\dot{M}(R_{D})}{2\pi
R(\Sigma_{a}+\Sigma_{d})}\,.\label{eq:vd}
\end{equation}
where $\dot{M}$ is as equation \ref{eq:layer4}. If we insert
$\dot{M}$ we can derive v$_{D}$$\propto 1/(\Sigma_{a}+\Sigma_{d})$.
Considering $\Sigma_{a}$+$\Sigma_{d}$ is larger at smaller radii,
the speed of R$_{D}$ decrease with time when it is moving inwards.
Finally, when R$_{D}$ moves to the very inner radii where
$\Sigma_{d} \gg \Sigma_{a}$, R$_{D}$ is almost halted.

\subsection{A.2 Outer pure MRI active disk beyond R$_{D}$} Beyond
R$_{D}$ the disk is purely MRI active with a constant $\alpha_{M}$.
Its surface density evolution can be solved by
\begin{equation}
\frac{\partial \Sigma}{\partial
t}=\frac{1}{R}\frac{\partial}{\partial R}\left[3
R^{1/2}\frac{\partial}{\partial R}\left(\nu\Sigma
R^{1/2}\right)\right]\,.\label{eq:evolve}
\end{equation}
with the boundary condition $\Sigma_{R_{D}}$=$\Sigma_{A}$. If $\nu=k
R^{n}$, this solution can be simplified by dividing  it into two
parts
\begin{equation}
\Sigma(R,t)=\Sigma'(R,t)+\Sigma_{A}\left(\frac{R}{R_{D}(t)}\right)^{-n-1/2}\,,\label{eq:sigmaprime}
\end{equation}
where the first term on the right comes from the disk evolution with
a zero surface density boundary condition at R$_{D}$
($\Sigma_{R_{D}}$=0), and the second term is the effect of the
non-zero boundary surface density at R$_{D}$. The first term behaves
like the similarity solution with a R$^{-1}$ part and a power law
decrease part at larger radius, and it eventually decreases to zero
with time. The surface density evolution is determined by the
competition between the first and the second term. During the infall
phase, the disk is massive, and the first term is larger than the
second term, thus the disk behaves like a similarity solution. Then
after infall stops, as the disk accretes and R$_{D}$ moves inwards,
both of these two terms decrease, and as shown above, R$_{D}$ moves
inwards more slowly. Eventually R$_{D}$ can be considered
independent of t; the first term goes to zero and the second term
dominates so that we have $\Sigma\propto R^{-1.5}$ (if n=1), which
distinguishes it from the self-similarity solution.

Next, we consider the late stage when R$_{D}$ is independent of time
analytically by the Green's function method. If the viscosity $\nu$
is a function of radius, equation \ref{eq:evolve} is a linear
equation for $\Sigma$ and can be solved by a Green's function (Lust
1952; Lynden-Bell \& Pringle 1974). If $\nu=k R^{n}$, the above
equation becomes
\begin{equation}
\frac{\partial \Sigma}{\partial
t}=\frac{1}{R}\frac{\partial}{\partial R}\left[3
R^{1/2}\frac{\partial}{\partial R}\left(k \Sigma
R^{n+1/2}\right)\right]\,,\label{eq:evolve2}
\end{equation}
and the radial dependence of $\Sigma$ is a linear combination of the
Bessel functions $J_{+\mu}$ and $J_{-\mu}$, where
$\mu^{2}=1/4(2-n)^{2}$ (Lynden-Bell \& Pringle 1974). However for a
fixed $\Sigma$ inner boundary condition at $R_{D}$,
\begin{equation}
\Sigma(R=R_{D}, t)=\Sigma_{A}\,,\label{eq:bound}
\end{equation}
we can substitute $\Sigma$ with $\Sigma'$
\begin{equation}
\Sigma(R,t)=\Sigma'(R,t)+\Sigma_{A}\left(\frac{R}{R_{D}}\right)^{-n-1/2}\,,\label{eq:sigmaprime2}
\end{equation}
so that equation \ref{eq:evolve2} and the boundary condition
\ref{eq:bound} changes to
\begin{equation}
\frac{\partial \Sigma'}{\partial
t}=\frac{1}{R}\frac{\partial}{\partial R}\left[3
R^{1/2}\frac{\partial}{\partial R}\left(k \Sigma'
R^{n+1/2}\right)\right]\,,\label{eq:evolve3}
\end{equation}
and
\begin{equation}
\Sigma'(R=R_{D},t)=0\,,
\end{equation}
which becomes a normal disk evolution equation of $\Sigma'$ with a
zero surface density boundary condition. The solution of $\Sigma'$
is well studied and this disk expands with limit
$t\rightarrow\infty$, $\Sigma'\rightarrow 0$, thus from equation
\ref{eq:sigmaprime2}, we derive $t\rightarrow\infty$,
$\Sigma\rightarrow \Sigma_{A}(R/R_{D})^{-n-1/2}$. Thus as long as
the disk evolves long enough the impact of any initial condition
will be washed out, and the boundary term dominates the surface
density distribution with $\Sigma\propto R^{-n-1/2}$.

In the irradiation dominated case, n=1, we transform equation
\ref{eq:evolve3} by writing $x=R^{1/2}$ and $\sigma=\Sigma'R^{3/2}$
(Pringle 1991) to get
\begin{equation}
\frac{\partial \sigma}{\partial t}=\frac{3
k}{4}\frac{\partial^{2}\sigma}{\partial x^{2}}\,.
\end{equation}
with the boundary condition $\sigma=0$ at $x=x_{D}=R_{D}^{1/2}$. The
general solution is then
\begin{equation}
\sigma(x,t)=\int_{-\infty}^{\infty}A_{\lambda}e^{-\lambda^{2}t}sin[\lambda(x-x_{D})/c]d\lambda\,.
\end{equation}
where $c^{2}=3k/4$. $A_{\lambda}$ is determined by the initial
conditions. Following Pringle (1991), in order to obtain the Green's
function, we set the initial condition with
\begin{equation}
\sigma(x,t=0)=\sigma_{0}\delta(x-x_{1})\,.\label{eq:ini}
\end{equation}
With the delta function Fourier transform $\int exp(2\pi x\lambda
i)d\lambda=\delta(x)$ and $x>x_{D}$, we derive
\begin{equation}
A_{\lambda}=-\frac{\sigma_{0}}{\pi c}sin[\lambda(x_{D}-x_{1})/c]\,.
\end{equation}
Thus the solution for the initial condition \ref{eq:ini} is
\begin{eqnarray}
\sigma(x,x_{1},t)&=&\frac{\sigma_{0}t^{-1/2}}{2\pi^{1/2}c}\{exp[-(x-x_{1})^{2}/3kt]\nonumber\\
&&-exp[-(x+x_{1}-2x_{D})^{2}/3kt]\}\,,
\end{eqnarray}
Finally, the solution $\sigma(x,t)$ for any initial condition
$\sigma(x,t=0)=\sigma'(x)$ is
\begin{eqnarray}
\sigma(x,t)&=&\int\frac{\sigma'(x_{1})t^{-1/2}}{2\pi^{1/2}c}\{exp[-(x-x_{1})^{2}/3kt]\nonumber\\
&&-exp[-(x+x_{1}-2x_{D})^{2}/3kt]\}dx_{1}\,.
\end{eqnarray}
and
\begin{equation}
\Sigma(R,t)=\sigma(R,t)R^{-3/2}+\Sigma_{A}\left(\frac{R}{R_{D}}\right)^{-3/2}\,,
\end{equation}
where the first term is the disk evolution with a zero boundary
condition and the second term is the fixed $\Sigma$ boundary effect.
As $t\rightarrow\infty$, $\Sigma\rightarrow
\Sigma_{A}(R/R_{D})^{-3/2}$.

Figure \ref{fig:outerdisk} shows the evolution of the constant
$\alpha=0.01$ disks with two different boundary conditions:
$\Sigma$(10 AU)=50 g cm$^{-2}$ (black curves) and $\Sigma$(10 AU)=0
g cm$^{-2}$ (red curves). The initial conditions are set as
\begin{equation}
\Sigma(R\leq50 \rm{AU},t=0)=50 \rm{g}\, \rm{cm}^{-2}\,,
\end{equation}
\begin{equation}
\Sigma(R>50 \rm{AU},t=0)=0 \rm{g}\,\rm{cm}^{-2}\,.
\end{equation}
As shown in figure \ref{fig:outerdisk}, the disk with zero boundary
condition behaves quite similar to the similarity solution. However,
for the $\Sigma$(10 AU)=50 g cm$^{-2}$ disk, the influence of the
boundary becomes significant with a part $\Sigma\propto R^{-1.5}$
and a outer part decreasing exponentially.

The mass accretion rate at R$_{D}$ decreases to zero eventually
($\Sigma\propto R^{-n-1/2}$). This can be shown by assuming $\Sigma=
g R^{m}$ and $\nu=k R^{n}$. Inserting these into
\begin{equation}
\dot{M}=6\pi R^{1/2}\frac{\partial}{\partial R}(\nu\Sigma
R^{1/2})\,,
\end{equation}
we find
\begin{equation}
\dot{M}=6\pi k g(m+n+\frac{1}{2})R^{m+n}\,.
\end{equation}
In the asymptotic case as discussed above with $\Sigma\propto
R^{-n-1/2}$, $\dot{M}=0$, which means at $t\rightarrow\infty$,
$\dot{M}\rightarrow 0$ at the inner boundary R$_{D}$, so no mass
flow in the disk. This can be simply understood because the disk has zero
torque with $\Sigma\propto R^{-n-1/2}$ if $\nu=k R^{n}$.

However, notice that this $\Sigma\propto R^{-3/2}$
behavior is only observed for layered disks with small or
negligible dead zone viscosity, so that at very late stage R$_{D}$ is
very small and almost fixed.

\subsection{A.3 Residual dead zone viscosity} If the residual dead zone
viscosity $\alpha_{rd}\gtrsim$10$^{-4}$, the dead zone can transport
the mass at a rate comparable to the accretion rate of the active
layer, considering $\dot{M_{i}}\propto\Sigma_{i}\alpha_{i}$ and
$\Sigma_{d}>>\Sigma_{a}$. If we assume the disk has a constant mass
accretion rate after a long period of time, then $\dot{M}$=-2$\pi
$R$\Sigma_{a}v_{Ra}$-2$\pi $R$\Sigma_{d}v_{Rd}$ is a constant and we
can integrate the angular momentum equation to derive
\begin{equation}
(\Sigma_{a}v_{Ra}+\Sigma_{d}v_{Rd})R^{3}\Omega=(\nu_{a}\Sigma_{a}+\nu_{d}\Sigma_{d})R^{3}\frac{d\Omega}{dR}+C\,.
\end{equation}
If we assume at the stellar surface $R_{*}$ $d\Omega/dR$ is 0, we
find C=-$\dot{M}\Omega$ R$_{*}^{2}$/2$\pi$. Thus
\begin{equation}
\nu_{a}\Sigma_{a}+\nu_{d}\Sigma_{d}=\frac{\dot{M}}{3\pi}\left[1-\left(\frac{R_{*}}{R}\right)^{1/2}\right]\,.
\end{equation}
Furthermore, if the disk is irradiation-dominated so that the temperature is
$\propto$R$^{-1/2}$, we have $\nu_{a}$=k$_{a}$R and
$\nu_{d}$=k$_{d}$R where k$_{a}$ and k$_{d}$ are constants and
proportional to $\alpha_{a}$ and $\alpha_{d}$. So
\begin{equation}
\Sigma_{d}=\frac{\dot{M}}{3\pi k_{d} R}
\left[1-\left(\frac{R_{*}}{R}\right)^{1/2}\right]-\frac{k_{a}}{k_{d}}\Sigma_{a}\,.
\end{equation}
If the dead zone transports mass at a rate higher than the active
layer, the last term can be neglected for order of magnitude
argument. And compared with a constant-$\alpha$ disk having the same
mass accretion rate, the layered disk has a surface density higher
than the constant-$\alpha$ disk by a factor of
$\alpha$/$\alpha_{d}$. If we assume the outer pure MRI active disk
and the inner layered disk has the same mass accretion rate, and the
outer disk only has $\nu_{a}$ while the inner dead zone is dominated
by $\nu_{d}$, the disk's surface density will increase by a factor
of $\alpha_{a}$/$\alpha_{d}$ within the dead zone outer radius
R$_{D}$, which is shown in Figure \ref{fig:deadzonevis2}. Also this
density change at R$_{D}$ is more gradual (equation \ref{eq:sigmad}
is a smooth function of R) than the zero dead zone viscosity case.

\section{B. GI disk fragmentation radius} Gammie (2001) has pointed
out that when the disk cooling timescale $t_{cool}\leq 3\Omega^{-1}$
the disk will fragment. By assuming local dissipation, Gammie (2001)
has shown that
\begin{equation}
t_{cool}=\frac{4}{9\gamma(\gamma-1)\alpha\Omega}\,.\label{eq:gammie}
\end{equation}
Thus, the disk will fragment if $\alpha>0.06$ (Rice et al. 2005).
However, the above fragmentation condition has only been tested for
disks without any irradiation. With irradiation dominated disk,
$t_{cool}$ is hard to be defined and $\alpha>$1 condition are used
 by Rafikov (2009) instead (Here we use $\alpha>$0.06 to be consistent
with the non-irradiated case.).

For a constant accretion rate we have
\begin{equation}
\alpha\frac{c_{s}^{2}}{\Omega}\Sigma=\frac{\dot{M}}{3\pi}\,.
\end{equation}
Combined with Q=1.5, we derive
\begin{equation}
\alpha=\frac{\dot{M}G}{2c_{s}^{3}}\,.\label{eq:alpha}
\end{equation}
Thus the disk will fragment if $\dot{M}$G/2c$_{s}^{3}$$>$0.06.

\begin{figure}
\epsscale{.80} \plotone{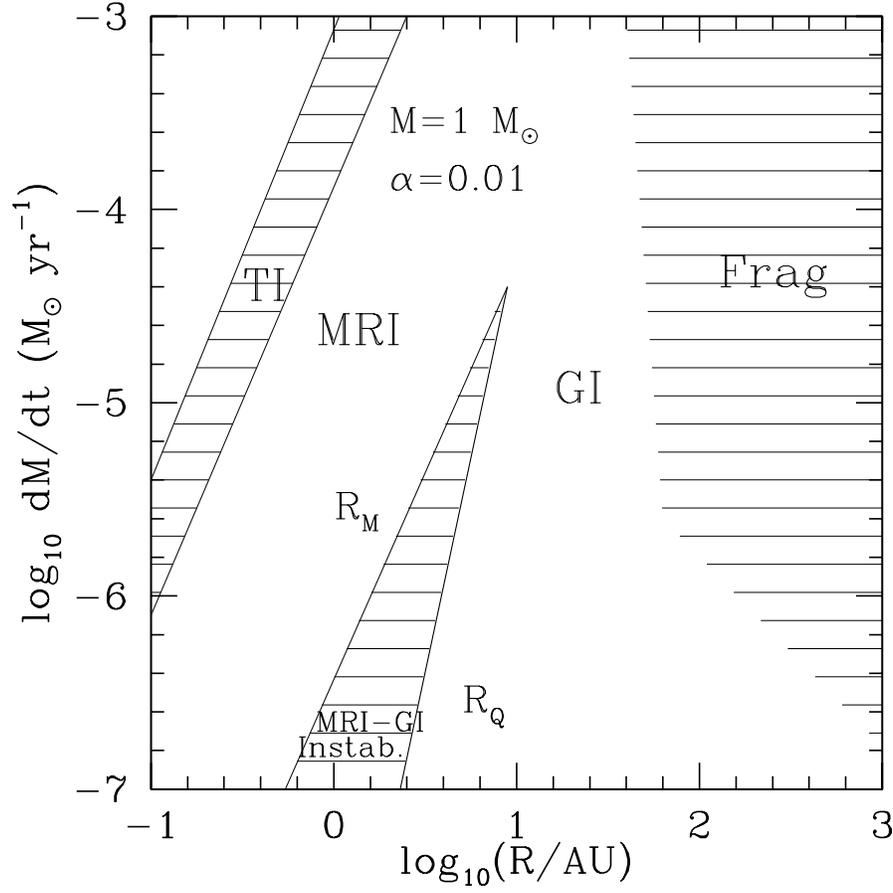} \caption{ Unstable regions in the
$R-\dot{M}$ plane for an accretion disk around $1 \msun$ central
star. The shaded region in the upper left shows the region subject
to classical thermal instability. The shaded region in the middle
shows where the central temperature of steady GI models exceeds an
assumed MRI trigger temperature of $1400$ K, thus subject to MRI-GI
instability. The lines labeled $R_M$ and $R_Q$ are the maximum MRI
and minimum GI steady accretion radius (the boundaries of the shaded
region, detail in Zhu et al. 2009b). The shaded region on the right
shows where the GI disk will fragment (See Appendix for details).
The fragmentation limit agrees with Cossins \etal (2009)b.}
\label{fig:m1}
\end{figure}

\begin{figure}
\epsscale{.80} \plotone{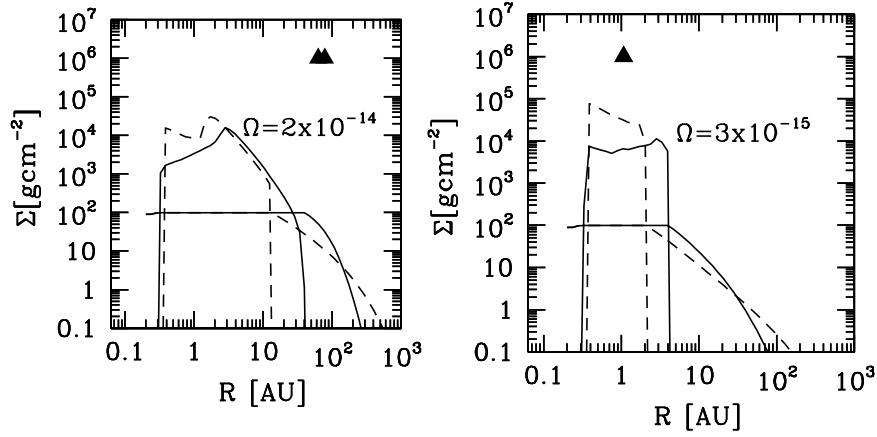} \caption{The disk surface density
with radii for different rotation cores as shown in figure
\ref{fig:highlowomega} at 0.3 million years (solid curves) and 1
million years (dashed curves).} \label{fig:difsurf}
\end{figure}

\begin{figure}
\epsscale{.80} \plotone{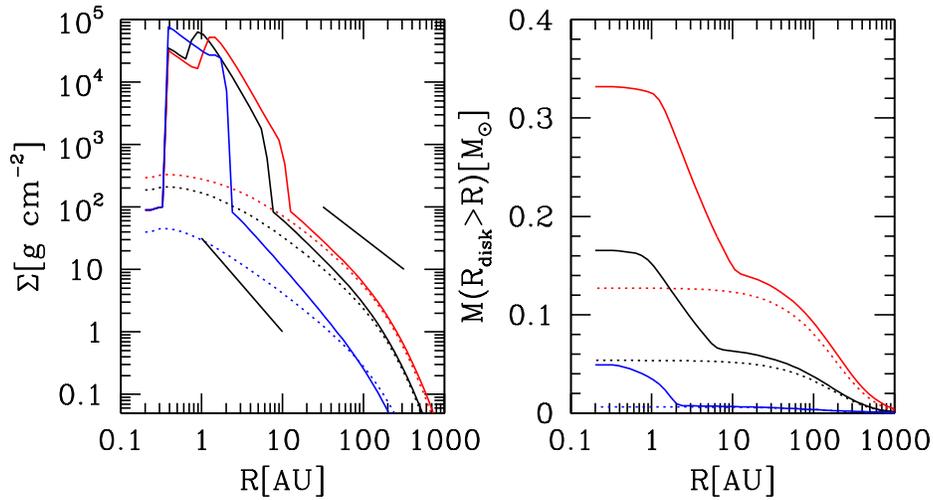} \caption{The disk surface density
distributions at 1 million years for layered models (solid curves)
and constant-$\alpha$ models ($\alpha$=0.01, dotted curves) for
$\Omega=2\times10^{-14}$ rad s$^{-1}$(red curves),
$\Omega=10^{-14}$rad s$^{-1}$ (black curves), and
$\Omega=3\times10^{-15}$rad s$^{-1}$ (blue curves). The two black
lines indicate $\Sigma\propto R^{-1}$ and $\Sigma\propto R^{-1.5}$
respectively. The right panels show the disks' mass beyond R
(M(R)=$\int_{R}^{\infty}$2$\pi$R$\Sigma$dR). } \label{fig:difmass}
\end{figure}

\begin{figure}
\epsscale{.80} \plotone{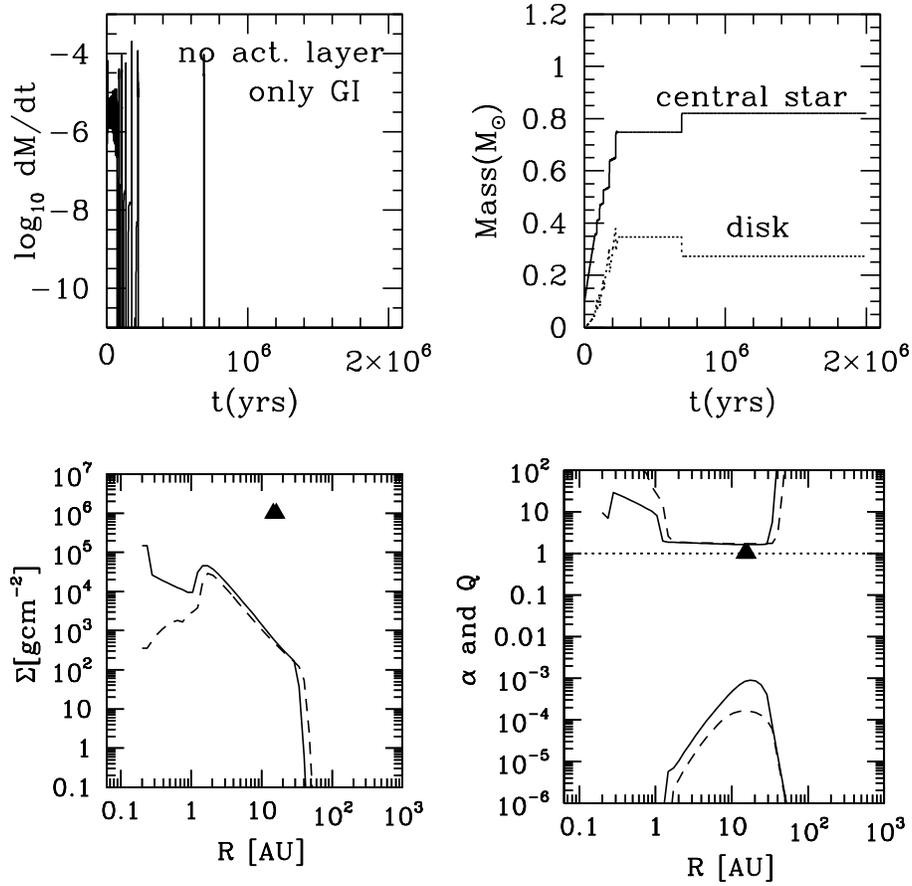} \caption{Disk evolution of the
model without the active layer, in which case only GI has been
considered to transport mass from the outer disk to the inner disk.
Other parameters are the same as the fiducial model. The upper left
panel shows the disk mass accretion rates and the upper right panel
shows the masses of the central star and disk with time. The lower
left panel shows the disk surface density distribution at 0.3
million years (solid curve) and 1 million years (dashed curve). The
lower right panel shows the disk $\alpha$ and Q parameter at 0.3 and
1 Myr. } \label{fig:deadzonevis}
\end{figure}

\begin{figure}
\epsscale{.80} \plotone{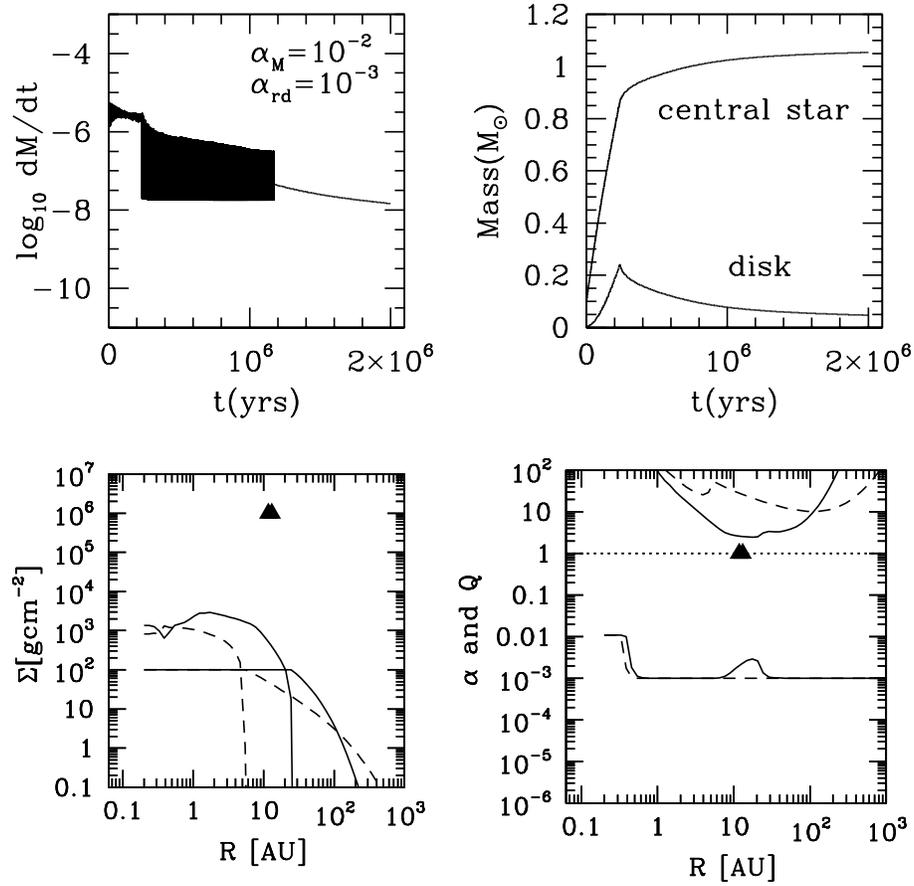} \caption{Disk evolution of the
model similar to our fiducial layered model but with a residual dead
zone viscosity ($\alpha_{rd}=10^{-3}$). The upper left panel shows
the disk mass accretion rates and the upper right panel shows the
masses of the central star and disk with time. The lower left panel
shows the disk surface density at 0.3 million years (solid curve)
and 1 million years (dashed curve). The lower right panel shows the
disk $\alpha$ and Q parameter at 0.3 and 1 million years.}
\label{fig:deadzonevis2}
\end{figure}

\begin{figure}
\epsscale{.80} \plotone{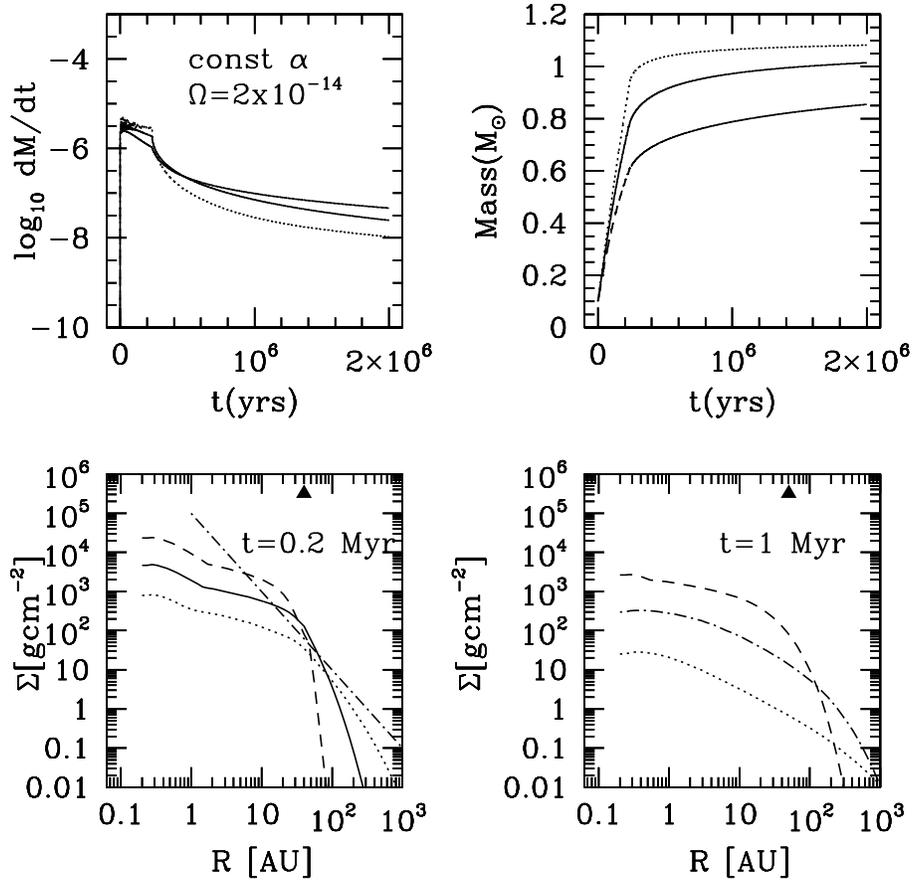} \caption{For constant-$\alpha$
disk (long dashed curve $\alpha$=10$^{-3}$;  solid curve
$\alpha$=0.01; dotted curve $\alpha$=0.1) with
$\Omega_{c}$=2$\times$10$^{-14}$ rotating core, upper panels show
the disk mass accretion rate (upper left panel) and central star
mass (upper right panel) with time, while lower panels show the disk
surface densities at 0.2 million years (lower left panel) and 1
million years (lower right panel) . The dash-dotted line in the
lower left panel shows the surface density corresponding to Q=1
disk.} \label{fig:constalphadm}
\end{figure}

\begin{figure}
\epsscale{.80} \plotone{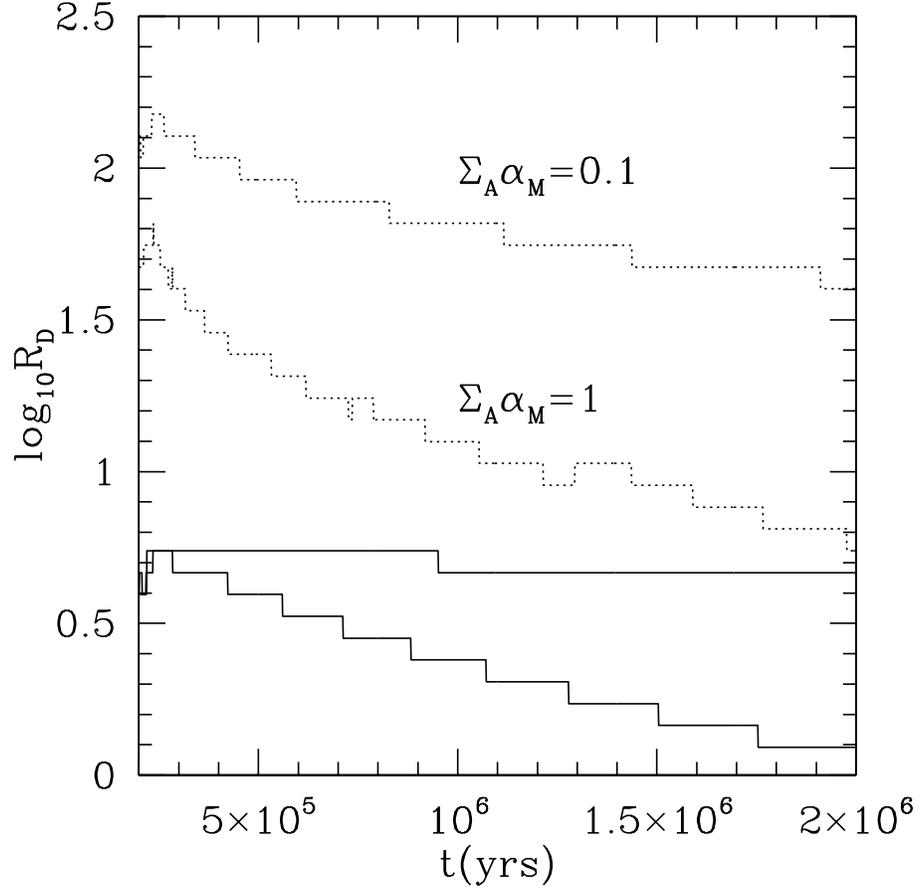} \caption{ In the layered disk
model, the dead zone outer radius R$_{D}$ with time for
$\Omega_{c}=2\times10^{-14}$rad s$^{-1}$ (upper two curves) and
$\Omega_{c}=3\times10^{-15}$rad s$^{-1}$ (lower two curves). Two
different configurations of the layered disk have been considered
($\Sigma_{A}\alpha_{M}$=1 and $\Sigma_{A}\alpha_{M}$=0.1). }
\label{fig:tran}
\end{figure}

\begin{figure}
\epsscale{.80} \plotone{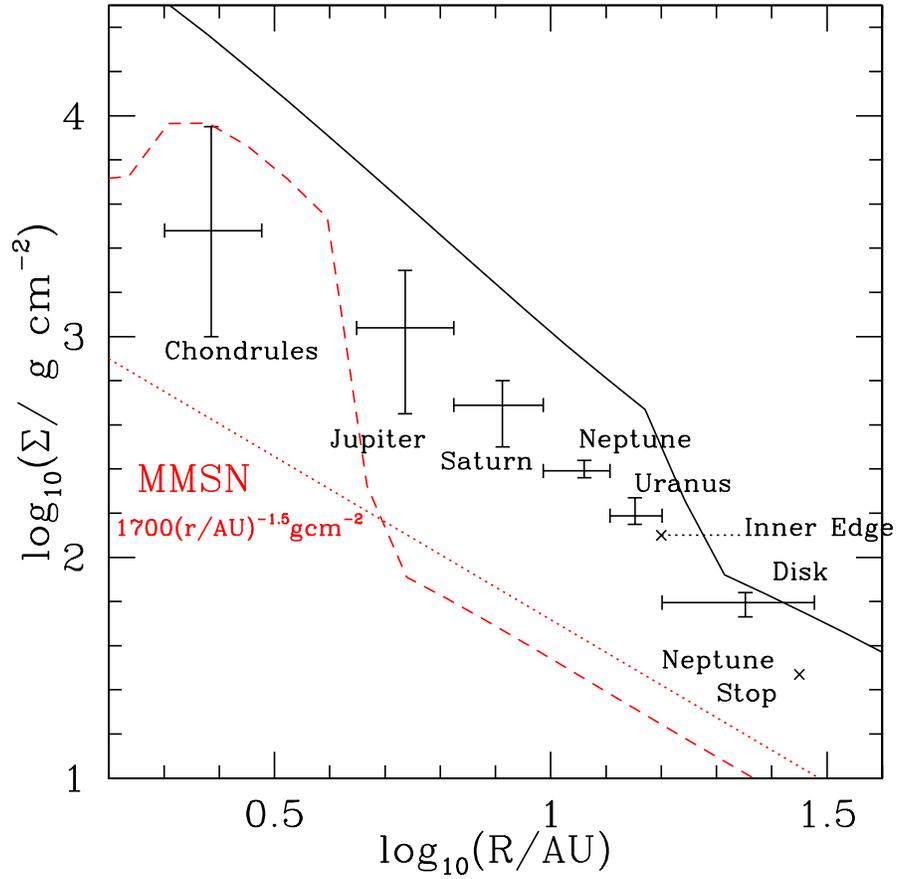} \caption{Our simulated disk
surface density compared with different MMSN models. The black solid
curve is the surface density from one of our layered models
($\Omega=2\times10^{-14}$rad s$^{-1}$ at 0.6 Myr) compared with the
surface density of the primordial disk constrained by Desch (2007).
The red dashed curve is the surface density from our
$\Omega=10^{-14}$rad s$^{-1}$ model at 2 Myr compared with the
Minimum mass solar nebulae from Weidenschilling (1977) (dotted red
curve). } \label{fig:desch}
\end{figure}

\begin{figure}
\epsscale{.80} \plotone{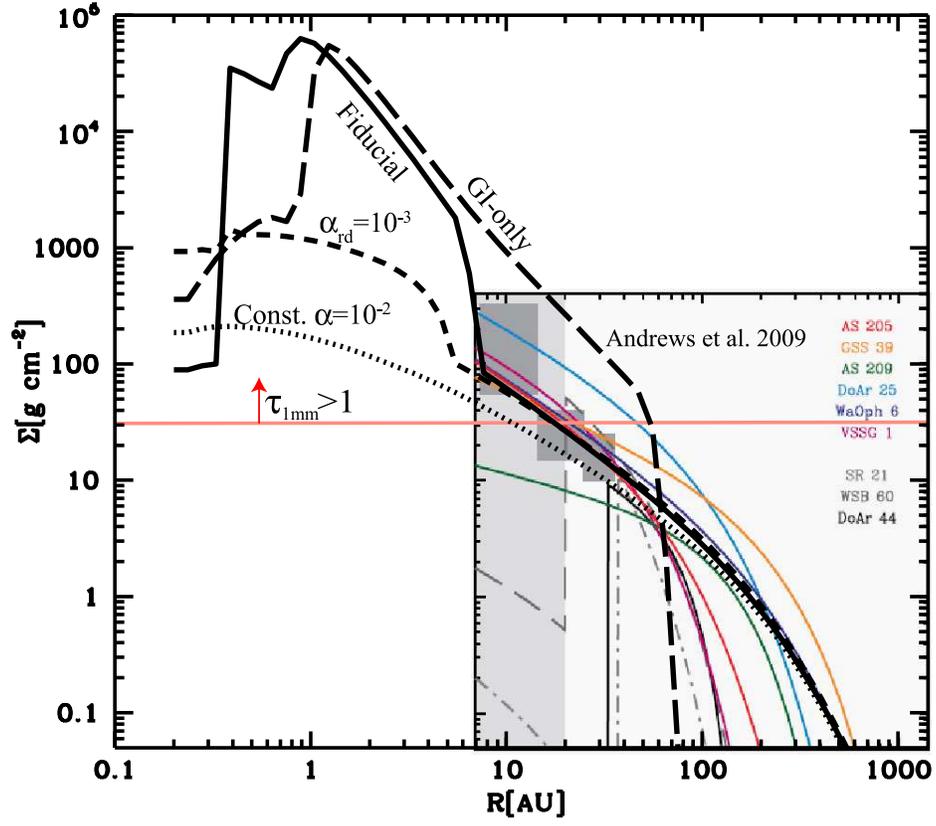} \caption{The disk surface density
distribution of our fiducial model (solid curve), fiducial model
with a residual dead zone viscosity ($\alpha_{rd}=10^{-3}$, dashed
curve), a constant-$\alpha$ ($\alpha$=0.01) model (dotted curve) and
a pure-GI model(long dashed curve) at 1 Myr compared with the disk
surface densities constrained by millimeter observations of
Ophiuchus (colored curves, Andrews et al. 2009).} \label{fig:submm}
\end{figure}

\begin{figure}
\epsscale{.80} \plotone{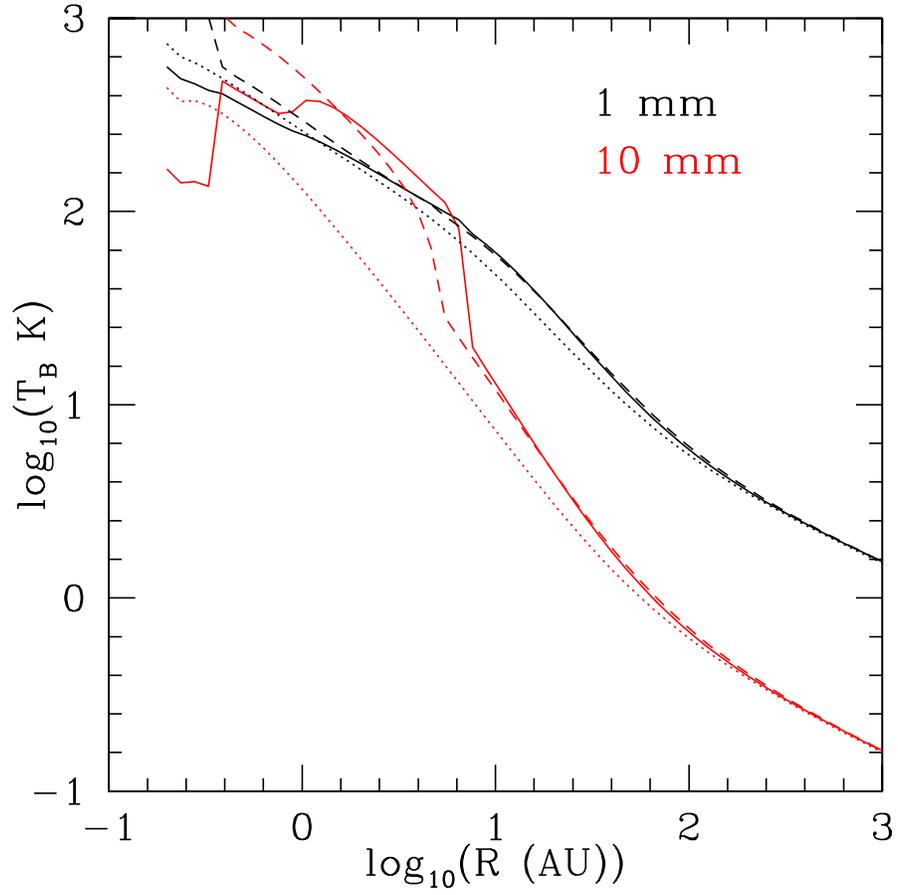} \caption{The disk surface
brightness temperature for the models in Figure \ref{fig:submm}: the
fiducial model (solid curve), the fiducial model with a dead zone
residual viscosity ($\alpha_{rd}=10^{-3}$) (dashed curve), and a
constant-$\alpha$ model with $\alpha$=0.01 (dotted curve) at 1
millimeter (black curves) and 10 millimeters (red curves). EVLA will
be able distinguish these different disk structures at 10
millimeters.} \label{fig:sb}
\end{figure}

\begin{figure}
\epsscale{.80} \plotone{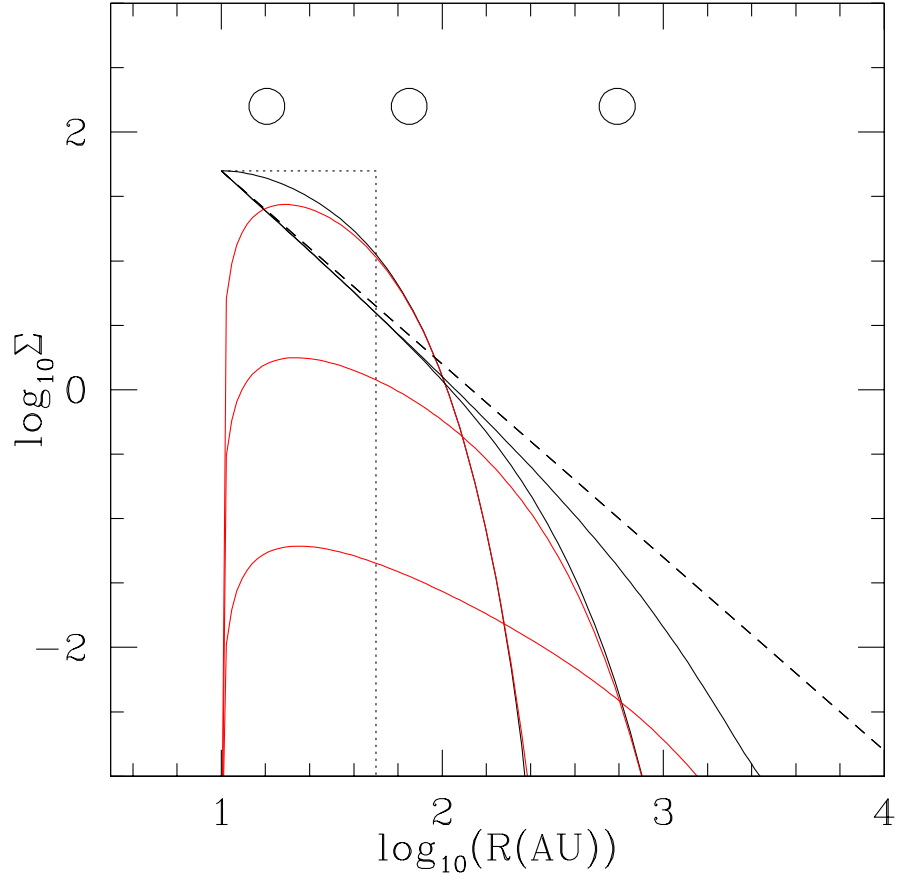} \caption{The viscous disk surface
density with radii at t=50, 500, 5000 kyr (curves from up to bottom)
with the fixed $\Sigma$ inner boundary condition (Black curves for
the inner boundary $\Sigma$=50 g cm$^{-2}$ while the red curves for
$\Sigma$=0 g cm$^{-2}$). The dotted curve extending to 50 AU shows
the disk initial condition, while the dashed line shows the
asymptotic limit $\rho\propto$R$^{-1.5}$. The open dots at the top
shows the scaling radius (see the definition in Lynden-Bell \&
Pringle 1974) R$_{t}$+10 AU at t=50, 500, 5000 kyr. }
\label{fig:outerdisk}
\end{figure}

For a viscous heating dominated Q=1 disk, at a given mass accretion
rate ($\dot{M}$), the relationship between T$_{c}$ and R is given in
Equation (23) in Paper I. If we reorganize the equation and assume
$\beta$=0 ($\kappa$=CT$^{\alpha}$P$^{\beta}$) with Q=3/2, we got
\begin{eqnarray}
T_{c}&=&3^{2/(7-2\alpha)}2^{-12/(7-2\alpha)}R^{-9/(7-2\alpha)}\left(\frac{\dot{M}}{\sigma}\right)^{2/(7-2\alpha)}\nonumber\\
&&C^{2/(7-2\alpha)}\left(\frac{\Re}{
\mu}\right)^{1/(7-2\alpha)}G^{1/(7-2\alpha)}M^{3/(7-2\alpha)}\pi^{-4/(7-2\alpha)}\,,
\end{eqnarray}
Inserting this into $\dot{M}$G/2c$_{s}^{3}$$>$0.06,
\begin{eqnarray}
R&>&0.5\times0.12^{(14-4\alpha)/27}\pi^{-4/9}G^{(-11+4\alpha)/27}C^{2/9}\sigma^{-2/9}\nonumber\\
&&\left(\frac{k}{\mu
m_{h}}\right)^{(8-2\alpha)/9}M^{1/3}\dot{M}^{(-8+4\alpha)/27}\,.
\end{eqnarray}
With our dust opacity $\alpha$=0.738 (Zhu \etal 2009 b), the
critical fragmentation radius R$\propto$$\dot{M}^{-0.19}$. Thus the
fragmentation radius is insensitive to $\dot{M}$ in the viscous
heating dominated case and $\sim$100 AU (Figure \ref{fig:m1}).

For an irradiation dominated case (low mass accretion rate), c$_{s}$
in equation \ref{eq:alpha} is determined by the irradiation
(equation \ref{eq:text}), thus
\begin{equation}
R>0.06\left( \dot{M}G\right)^{-4/3}\left(\frac{k}{\mu
m_{h}}\right)^{2}\left(\frac{f L}{4\pi\sigma}\right)^{1/2}\,.
\end{equation}
In this case the critical fragmentation R has a sharper dependence
on $\dot{M}^{-4/3}$ (Figure \ref{fig:m1}). The GI fragmentation
region is outlined in Figure \ref{fig:m1}, which agrees with Figure
10 of Cossins \etal (2009)b.

\clearpage

\clearpage

\clearpage

\clearpage

\clearpage
\begin{table}
\begin{center}
\caption{1D2Z models at 0.3 and 1 Myr\label{tab1}}
\begin{tabular}{ccccccc}

\tableline\tableline
cloud $\Omega$   & $\Sigma_{A}$  & $\alpha_{M}$  & R$_{D}$ 0.3/1 Myr & M$_{in}$\tablenotemark{a} & M$_{o}$\tablenotemark{b} & M$_{t}$\tablenotemark{c}\\
& g cm$^{-2}$ & & AU & $\msun$ & $\msun$ & $\msun$\\
\tableline
2$\times$10$^{-14}$ & 100 & 0.01 & 40/13 & 0.29/0.19 & 0.18/0.14  & 0.47/0.33   \\
2$\times$10$^{-14}$ & 10 & 0.01 & 127/66 & 0.42/0.33 & 0.05/0.09  & 0.48/0.42  \\
2$\times$10$^{-14}$ & 0 & 0.01 & -/- & 0.47/0.41 & -/-  & 0.47/0.41 \\
1$\times$10$^{-14}$ & 100 & 0.01 & 24/7.6 & 0.17/0.1 & 0.07/0.06  & 0.24/0.16   \\
1$\times$10$^{-14}$ & 10 & 0.01 & 47/34 & 0.32/0.23 & 0.008/0.022  & 0.32/0.26   \\
3$\times$10$^{-15}$ & 100 & 0.01 & 4.6/2.4 & 0.05/0.04 & 0.006/0.007  & 0.057/0.049   \\
3$\times$10$^{-15}$ & 10 & 0.01 & 5.5/4.6 & 0.06/0.06 & 0.0004/0.001  & 0.063/0.063   \\
3$\times$10$^{-15}$ & 0 & 0.01 & -/- & 0.064/0.064 & -/-  & 0.064/0.064   \\
 \tableline
\end{tabular}
\tablenotetext{a}{The disk mass within R$_{D}$}
\tablenotetext{b}{The disk mass beyond R$_{D}$}
\tablenotetext{c}{The total disk mass}

\end{center}
\end{table}
\clearpage

\FloatBarrier

\end{document}